\documentclass[article,nojss,shortnames,noheadings]{jss}

\usepackage{thumbpdf,lmodern}
\graphicspath{{Figures/}} 
\usepackage[utf8]{inputenc}

\usepackage{thumbpdf,lmodern}
\usepackage{amssymb,amsmath,amsthm,amsfonts,mathtools}
\usepackage{framed}
\usepackage{bm,graphicx,algorithmic,algorithm}
\usepackage[font={small}, labelfont={bf}]{caption}
\usepackage{color}
\definecolor{light-gray}{gray}{0.95}
\usepackage{listings}
\usepackage{caption}
\usepackage{subfig}
\usepackage{comment}
\usepackage{array}

\newcommand{\class}[1]{`\code{#1}'}

\newcommand{\Model}{\bm{\mathcal{M}}}

\newcommand{\data}{\bm{y}}
\newcommand{\dataObs}{\bm{y}_{0}}
\newcommand{\datasim}{\bm{y}_{\text{sim}}}
\newcommand{\distance}{\rho}
\newcommand{\prior}{\pi}

\newcommand\ith[1]{{#1}^{(i)}}
\newcommand\kth[1]{{#1}^{(k)}}

\newcommand{\speedup}[2]{\mathcal{S}_{#1}(#2)}
\newcommand{\eff}[2]{\mathcal{E}_{#1}(#2)}
\newcommand{\appropto}{\mathrel{\vcenter{
  \offinterlineskip\halign{\hfil$##$\cr
    \propto\cr\noalign{\kern2pt}\sim\cr\noalign{\kern-2pt}}}}}

\usepackage{tikz}

\pgfdeclarelayer{background}
\pgfdeclarelayer{foreground}
\pgfsetlayers{background,main,foreground}
\usetikzlibrary{arrows,shapes,positioning,decorations.markings}
\usetikzlibrary{decorations.markings}
\tikzset{element/.style={rectangle, 
                            minimum width= 3em,
                            minimum height = 2em,
                            thick,
                            draw = blue,
                            top  color=blue!20!white!20, bottom  color=blue!70!white,
                            }}    
\tikzset{element_elip/.style={ellipse, 
                            minimum width= 3em,
                            minimum height = 2em,
                            top  color=black!20!white!20, bottom  color=black!70!white,
                            }} 
\tikzset{element_circ/.style={circle, 
                            top  color=black!20!, bottom  color=black!70!,
                            }}      
\tikzstyle{fancytitle} =[fill=red, text=white]

\author{Ritabrata Dutta\\
  University of Warwick
\And Marcel Schoengens\\
  ETH Zurich
\And Lorenzo Pacchiardi\\
  University of Oxford\\
\AND Avinash Ummadisingu\\
  Universit\`a della Svizzera italiana \\
\And Nicole Widmer\\
  ETH Zurich \\
\And Pierre K\"unzli\\
  University of Geneva \\
\AND Jukka-Pekka Onnela\\
  Harvard University \\
\And Antonietta Mira\\
  Universit\`a della Svizzera italiana \\
  Universit\`a dell'Insubria
}
\Plainauthor{Ritabrata Dutta, Marcel Schoengens, Lorenzo Pacchiardi,
    Avinash Ummadisingu, Nicole Widmer, Jukka-Pekka Onnela, Antonietta Mira}

\title{\pkg{ABCpy}: A High-Performance Computing Perspective to 
      Approximate Bayesian Computation}
\Plaintitle{ABCpy: A High-Performance Computing Perspective to 
            Approximate Bayesian Computation}
\Shorttitle{\pkg{ABCpy}: A High-Performance Computing Perspective to 
      ABC}

\Abstract{
  \pkg{ABCpy} is a highly modular scientific library for approximate
  Bayesian computation (ABC) written in \proglang{Python}.  The main
  contribution of this paper is to document a software engineering effort
  that enables domain scientists to easily apply ABC to their research
  without being ABC experts; using \pkg{ABCpy} they can easily run large
  parallel simulations without much knowledge about parallelization. 
  Further, \pkg{ABCpy} enables ABC experts to easily develop new inference
  schemes and evaluate them in a standardized environment and to extend the
  library with new algorithms.  These benefits come mainly from the
  modularity of \pkg{ABCpy}.  We give an overview of the design of
  \pkg{ABCpy} and provide a performance evaluation concentrating on
  parallelization.  This points us towards the inherent imbalance in some of
  the ABC algorithms.  We develop a dynamic scheduling \pkg{MPI} implementation to
  mitigate this issue and evaluate the various ABC algorithms according to
  their adaptability towards high-performance computing.
}

\Keywords{ABC, HPC, \pkg{Spark}, \pkg{MPI}, parallel, imbalance, \proglang{Python} library}
\Plainkeywords{ABC, HPC, Spark, MPI, parallel, imbalance, Python library}

\Address{
  Ritabrata Dutta\\
  Department of Statistics\\
  University of Warwick, United Kingdom\\
  E-mail:~\email{Ritabrata.Dutta@warwick.ac.uk}}

\begin{document}

\section{Introduction} \label{sec:Introduction}

Today, computers are used to simulate different aspects of nature.  Natural
scientists traditionally hypothesize models underlying natural phenomena. 
As a running example throughout the paper, we will consider a popular
weather prediction benchmark model known as the Lorenz95 model
\citep{Lorenz_1995}, which represents an idealized weather system with two
sets of variables, the former evolving slowly in time and the latter
evolving much faster.  The evolution follows a set of differential
equations, and each of the slow variables is coupled to three neighbor ones
and to a subset of the fast variables (that outnumber the slow ones), and
similarly for the evolution of the fast variables.  We will focus on a
stochastic modification of the original model due to \cite{Wilks_2005}, in
which the fast variables are unobserved and their effect on the slow
variables is replaced by a stochastic forcing term; see Appendix
\ref{app:lorenz} for more details.  The implementation of the model is a
discrete time integration of the set of stochastic differential equations,
each integration of the model corresponding to a possible trajectory with a
finite timestep.  The equations depend on a set of parameters collectively
called $ \theta $, on which we want to perform inference given an
observation.  Therefore, denoting the model by $\Model$ and the observed
slow variables at timestep $ t $ by $y^{(t)}$, an integration of the model
yields:
\begin{eqnarray}
\label{eq:simulator}
\Model (y^{(0)}, \theta) \rightarrow \lbrace y^{(t)}, \ t=1, \ldots,  T \rbrace,
\end{eqnarray}
where the initial configuration $y^{(0)}$ is assumed to be known. 
%
%
Simulator-based models as the above one\footnote{In this manuscript, we will
use the term simulator-based model to refer to a model that enables direct
simulation of model outcomes using a set of stochastic rules.  This term is
well established within the ABC literature, but we point out that these
types of models are sometimes called mechanistic models or agent based
models in different fields of science.} are used in a wide range of
scientific disciplines to simulate different aspects of nature, ranging from
dynamics of sub-atomic particles \citep{Martinez_2016} to evolution of human
societies \citep{Turchin_2013} and formation of universes
\citep{Schaye_2015}.

However, often the true parameter $\theta^0$ of simulator-based models is
not known.  If the true parameter value could be learned rigorously in a
data-driven manner, we could substantially improve the accuracy of these
models.  Consider the problem of estimating the true value and quantifying
uncertainty in $\theta$ based on an observed dataset $\dataObs$, e.g.,~in
the Lorenz95 model $\dataObs \equiv \lbrace y^{(t)}_0, \ t=1, \ldots, T
\rbrace$.  A further extension of this inferential problem is the selection
of a model, given an observed dataset, from a set of possible models. 
Traditional methods in statistics can infer, from the observed data, model
and corresponding parameters and quantify the associated uncertainty only
when {the data generating mechanism has a known likelihood function.} {In
many cases, however, we may not have access to an explicit formula for the
latter or, if we have, its evaluation can be too computationally expensive;
for instance, if the data generating model consists of the integration of a
set of stochastic differential equations (as in the Lorenz95 model above),
there is no easy way to evaluate the likelihood of each integration for a
set of parameter values.  Alternatively, it can be that the likelihood
depends on the inversion of a high-dimensional covariance matrix, which can
be very costly.}

{In the above scenarios}, approximate Bayesian computation (ABC)
\citep{Tavare1997, Pritchard1999, Beaumont2002} can still offer a way to
perform sound statistical inference, e.g.,~point estimation, hypothesis
testing, and model selection.  ABC methods infer parameters by first
simulating a dataset using a proposed parameter value and accepting or
rejecting that parameter value either by comparing the closeness of the
simulated dataset to the observed dataset, usually through the use of
summary statistics, or by approximating the likelihood function using
simulated datasets \citep{Wood_2010, thomas2016likelihood}.  We direct
interested readers to the review paper by \citet{Lintusaari_2016}.

The necessity to simulate datasets from simulator-based models makes ABC
algorithms extremely expensive when this forward simulation itself is
costly.  Applications of ABC algorithms to complex problems show the
necessity of adapting them to high-performance computing (HPC) facilities
and developing an ecosystem where new ABC algorithms can be investigated
while respecting the architecture of existing computing facilities.  ABC and
HPC were first brought together in the \pkg{ABC-sysbio} package of
\cite{Liepe_2010} for the systems biology community, where the sequential
Monte Carlo ABC (SMCABC) algorithm \citep{Toni_2009} was efficiently
parallelized using graphics processing units (GPUs).

Our goal is to overcome the need for users to have knowledge of parallel
programming, as is required for using \pkg{ABC-sysbio}, and also to make a
software package available for scientists across domains.  These objectives
were partly addressed by parallelization of SMCABC using
\pkg{MPI}~/~\pkg{OpenMPI} \citep{Stram_2015}, and by making SMCABC available for the
astronomical community \citep{Jennings_2016}.  Regardless of these advances,
a recent ABC review article \citep{Lintusaari_2016} highlights the depth and
breadth of available ABC algorithms, which can be made efficient via
parallelization using an HPC environment
\citep{Kulakova_2016,Chiachio_2014}.  These developments emphasize the need
of a generalized HPC supported platform for efficient ABC algorithms, which
can be parallelized on multi-processor computers or computing clusters and
is accessible to a broad range of scientists.

We address the need for a user-friendly scientific library for ABC
algorithms by introducing \pkg{ABCpy}, which is written in \proglang{Python}
\citep{python} and designed in a highly modular fashion.  Most existing ABC software suites
are mainly domain-specific and optimized for a narrower class of problems. 
Our main goal was to make \pkg{ABCpy} modular, which makes it intuitive to
use and easy to extend.  Further, it enables users to run ABC sampling
schemes in parallel without too much re-factoring of existing code. 
\pkg{ABCpy} includes likelihood free inference schemes, both based on
discrepancy measures and approximate likelihood, providing a complete
environment to develop new ABC algorithms. The source code can be downloaded
from \url{https://github.com/eth-cscs/abcpy}.

For parallelization of ABC algorithms, we use the map-reduce paradigm.  This
choice was motivated by our experience that ABC algorithms are usually
parallelizable in a loosely coupled fashion.  Additionally, opting for
map-reduce we were able to implement parallelization backends in two
different frameworks (namely, \pkg{Apache Spark} \citep{ApacheSpark} and \pkg{MPI}
\citep{message2012mpi}), that target the needs of two different but
important communities (correspondingly, industry users and researchers). 
Thus, the choice of map-reduce increases the user's flexibility given widely
available commercial cloud computing facilities.  In
Section~\ref{sec:API_design} we discuss in detail the reasons for these
choices.

Of particular interest to practitioners might be the MPI backend since in
contrast to \pkg{Spark}, \pkg{MPI} is a low level communication framework without
sophisticated task scheduling facilities.  A straightforward \pkg{MPI}
implementation can therefore result in load imbalance between the different
workers for the ABC algorithms.  To handle this, we use a greedy approach to
dynamically allocate map tasks to workers in our \pkg{MPI} backend.  More details
on this can be found in Section~\ref{sec:dynamicMPI}.

We give a brief description of ABC (Section~\ref{sec:ABC}) and of the
structure of the software suite \pkg{ABCpy}
(Section~\ref{sec:ABCpy_sructure}) with a specific focus on modularity
(Section~\ref{sec:modular}) and parallelism.  Section~\ref{sec:Parallelism}
deals with the different map-reduce implementations available through
\pkg{ABCpy} and a detailed comparison of the speed-up and efficiency for ABC
algorithm using the Lorenz95 model; specifically, the scalability of
different ABC algorithms is compared in
Section~\ref{sec:imbalance_classify_abc}.  Finally, we compare our package
with similar ones in Section~\ref{sec:comparison_other_packages}, where we
also give a detailed overviews of the most important features that our
package implements that are not available in any other up to now, namely the
possibility of automatically learn summary statistics, the handling of
co-occurring datasets, the use of nested parallelization and the diagnostic
checks.  We conclude in Section~\ref{sec:conclusion} with some final
remarks.

\section{ABC}\label{sec:ABC}

We can quantify the uncertainty of the unknown parameter $\theta$ by a
posterior distribution $p(\theta \mid \data)$ given the observed dataset $\data =
\dataObs$.  A posterior distribution can be written, by Bayes' Theorem, as:
\begin{eqnarray} 
p(\theta \mid \data) = \frac{\prior(\theta)p(\data \mid \theta)}{m(\data)}, 
\end{eqnarray}
where $\prior(\theta)$, $p(\data \mid \theta)$ and $m(\data) = \int
\pi(\theta)p(\data \mid \theta)d\theta$ are, correspondingly, the prior
distribution on the parameter $\theta$, the likelihood function, and the
marginal likelihood.  The prior distribution $\pi(\theta)$ ensures a way to
leverage the learning of parameters with prior knowledge.  If the likelihood
function can be evaluated, at least up to a normalizing constant, then the
posterior distribution can be approximated by drawing a sample of parameter
values using (Markov chain) Monte Carlo sampling schemes \citep{Robert2005}. 
In {many} real-world problems, however, the analytic form of the posterior
distribution is unknown because the likelihood is not analytically
available.  This is typical for simulator-based models for which the
likelihood function is often intractable or difficult to compute (as for
instance the Lorenz model above or other integrations of stochastic differential
equation models), and
therefore the inference schemes are adapted following two alternative
approaches: (i) by measuring the discrepancy between simulated and observed
dataset, and (ii) by approximating the likelihood function.

\subsection{Measuring discrepancy} \label{sec:measuring_discrepancy}

In the simplest ABC implementation we forward simulate from the model,
$p(\data \mid \theta)$, producing a synthetic dataset $\datasim$ for a given
parameter value $\theta$, and measure the closeness between $\datasim$ and
$\dataObs$ using a pre-defined discrepancy function
$\distance(\datasim,\dataObs)$.  Based on this discrepancy measure, ABC
accepts the parameter value $\theta$ when $\distance(\datasim,\dataObs)$ is
less than a pre-specified threshold value $\epsilon$.  This simple algorithm
will be referred to as rejection ABC (RejectionABC).  A review of different
methods based on discrepancy can be found in \cite{Marin_2012} and
\cite{Lintusaari_2016}; {in Section~\ref{sec:ABCpy_alg}, we briefly describe
those which we implement in \pkg{ABCpy}.  }

To implement any ABC sampling scheme, we need to define how to measure the
discrepancy between $\datasim$ and $\dataObs$.  As the dataset can be of
varied type and complexity (e.g.,~high-dimensional time-series or network
data), in practice discrepancies are measured using informative summary
statistics extracted from the dataset.  We therefore need to define two
functions: one for computing the summary statistics from the dataset, and
one for measuring the discrepancy between them.  From now on, we will denote
these two functions as $statistics$ and $distance$, which need to be defined
by the user and are problem specific.

For illustration and comparison, in this paper we will consider the Lorenz95
model for numerical weather prediction \citep{Lorenz_1995, Wilks_2005} with
a stochastic modification, as discussed above.  For this model, a possible
choice of \emph{statistics} are the summary statistics suggested in
\cite{Hakkarainen_2012} called \code{HakkarainenLorenz} (details in
Appendix~\ref{app:lorenz}), while we can use as \emph{distance} the
\code{Euclidean} distance.  {Besides the latter, \pkg{ABCpy} also implements
distances based on logistic regression (\code{LogReg}) and penalized
logistic regression (\code{PenLogReg}) classifiers \citep{Gutmann_2014};
both work by fitting the classifier to distinguish between observed datasets
and datasets generated from the model with a fixed parameter value and by
using the resulting classification accuracy as discrepancy measure.}
Finally, we also provide a \code{Wasserstein} distance
\citep{peyre2019computational}.  Specifically, if several independent and
identically distributed (i.i.d.) observations are available, the latter can
be used as in \cite{bernton2019}, by generating i.i.d. simulations from the
model for each parameter value and afterwards computing the Wasserstein
distance between the empirical distributions defined by the observed and
synthetic dataset.
   
\subsection{Approximate likelihood}\label{sec:appr_lik}

The second approach is based on directly approximating the likelihood
function at $\theta$, up to a constant, using the data, $\datasim$,
simulated for that given parameter value $\theta$.  Following the
pseudo-marginal likelihood idea of \cite{Andrieu_2009}, an unbiased
approximation of the likelihood function can then be used in a traditional
Monte Carlo sampling scheme to sample from the posterior distribution.

Similarly to the scheme described in
Section~\ref{sec:measuring_discrepancy}, to perform any approximate
likelihood based sampling scheme we need to define two functions.  We
require the \emph{statistics} function and, additionally, we need a function
to compute the approximate likelihood based on the extracted summary
statistics from $\datasim$.  We denote this function by \code{approx\_lhd}
and the user needs to choose from one of the three currently available
implementations of \code{approx\_lhd} in \pkg{ABCpy}: 
\begin{itemize}
\item Synthetic likelihood (\code{SynLikelihood}) \citep{Wood_2010}, which works by
assuming the statistics to have a multivariate normal likelihood and by
estimating the mean and covariance parameters from $\datasim$.
\item Semiparametric Synthetic likelihood (\code{SemiParametricBSL}) \citep{an2020robust}, which is an extension of the above in which the likelihood of the summary statistics is represented as the product of a Gaussian copula (whose parameters are estimated from $\datasim$) and univariate marginals obtained with kernel density estimates.
\item Penalized logistic regression (\code{PenLogReg}) \citep{thomas2016likelihood}, which
	instead builds a likelihood approximation by fitting a probabilistic
	classifier between data generated from the model for a fixed parameter value
	and data generated from the marginal $ p(x) $; this in fact approximates the
	ratio $ \frac{p(x \mid \theta)}{p(x)} $, which is proportional to the likelihood
	with respect to $ \theta $.
\end{itemize}

\subsection{Implemented algorithms}\label{sec:ABCpy_alg}

In \pkg{ABCpy}, besides the standard RejectionABC algorithm, we implement
widely used and advanced variants, namely: population Monte Carlo ABC
(PMCABC) \citep{Beaumont2010, Toni_2009}, sequential Monte Carlo ABC
(SMCABC) \citep{del2012adaptive}, replenishment sequential Monte Carlo ABC
(RSMCABC) \citep{drovandi2011estimation}, adaptive population Monte Carlo
ABC (APMCABC) \citep{lenormand2013adaptive}, ABC with subset simulation
(ABCsubsim) \citep{Chiachio_2014}, and simulated annealing ABC (SABC)
\citep{Albert_2015}.  \pkg{ABCpy} also includes a parallelized version of a random
forest ensemble model selection algorithm \citep{Pudlo_2015}.

In Algorithm~\ref{alg:PMCABC}, we provide a description of the PMCABC
algorithm, which we will use in the following to illustrate the idea of ABC
algorithms and their parallelization.  In a nutshell, PMCABC considers a set
of sample points on the parameter space (particles) which evolve across
iterations.  At iteration $ t $, the position of each particle is perturbed
with a \emph{perturbation kernel}, and then simulations from the model are
run until the simulation matches the observation at some level of tolerance
$ \epsilon_t $, according to the considered distance and statistics.  The
sequence of thresholds $ \epsilon_1, \epsilon_2, \ldots $ need to be
decreasing to ensure convergence of the algorithm; they can be either fixed
a priori by the user or defined as some quantile of the distances at
previous iteration.
\begin{algorithm}[t!]
\caption{Population Monte Carlo ABC (PMCABC) algorithm for generating $N$
  samples from the approximate posterior distribution.  Here
  $K_t(\cdot \mid \theta, \Sigma_{t-1})$ is the perturbation kernel, and
  \code{weigthed-Covariance} (not shown here) updates the covariance matrix
  of the perturbation kernel according to the drawn samples and weights.}
\label{alg:PMCABC}
\begin{algorithmic}[1]
\REQUIRE Specify $q_{\epsilon} \in [0,100]$ and a decreasing sequence of
thresholds $\epsilon_1 \ge \epsilon_2 \ge \dots \ge \epsilon_T$ for $T$
iterations.
\FOR{$i=1$ \TO $N$}
\REPEAT
\STATE Generate $\theta$ from the prior $\prior(\cdot)$
\STATE Generate $\datasim$ from $\Model$ using $\theta$
\UNTIL {$\distance(\datasim,\dataObs) \le \epsilon_1$}
\STATE $d^{(i)} = \distance(\datasim,\dataObs)$
\STATE $\ith\theta_1 \leftarrow \theta$
\STATE $\ith\omega_1 \leftarrow 1/N$
\ENDFOR
\STATE $\Sigma_1 \leftarrow 2*\mbox{\code{weighted-Covariance}}(\theta_1,
\omega_1)$
\FOR{$t=2$ \TO $T$}
\STATE $\epsilon_t = \max(q_{\epsilon}$-th percentile of $d$, $\epsilon_t$) 
\FOR{$i=1$ \TO $N$}
\REPEAT
\STATE Draw $\theta^*$ from among $\theta_{t-1}$ with probabilities $\omega_{t-1}$
\STATE Generate $\theta$ from $K_t(\theta^*,\Sigma_{t-1})$
\STATE Generate $\datasim$ from $\Model$ using $\theta$
\UNTIL {$\distance(\datasim,\dataObs) \le \epsilon_t$}
\STATE $d^{(i)} = \distance(\datasim,\dataObs)$
\STATE $\ith\theta_t \leftarrow \theta$
\STATE $\ith\omega_t \leftarrow
\prior(\theta)/(\sum^N_{k=1}\kth\omega_{t-1}K_t(\theta \mid \kth\theta_{t-1},
\Sigma_{t-1}))$
\ENDFOR
\STATE Normalize $\ith\omega_t$ over $i = 1,\ldots, N$
\STATE $\Sigma_t \leftarrow 2*\mbox{\code{weighted-Covariance}}(\theta_t,
\omega_t)$
\ENDFOR
\end{algorithmic}
\end{algorithm}
Similarly, the other algorithms which \pkg{ABCpy} implements (except for
RejectionABC) follow the idea of considering a set of particles and evolving
them across iterations, but differ in how the particles are perturbed from
one iterations to the next (which is linked to drawing simulations from the
model) and how the importance weights are computed.  Some of the most recent
algorithms (such as SABC and APMCABC) usually provide faster convergence to
the posterior distribution and a smaller number of simulations required.
 
We can however classify the above algorithm into two groups, based on how
simulations from the model are run.  In one group, algorithms have an
explicit acceptance step similar to Lines~2--5 of PMCABC (see
Algorithm~\ref{alg:PMCABC}), where we keep simulating $\datasim$ until the
condition $\distance(\datasim, \dataObs)< \epsilon$ (for an adaptively
chosen threshold $\epsilon$), is met and the perturbed parameter is
accepted.  By enforcing this explicit acceptance for each perturbed
parameter, we have a theoretical warranty that the accepted parameters are
drawn from an approximate posterior distribution indexed by the chosen
threshold $\epsilon$.  For the second group of algorithms, we do not impose
explicit acceptance but we rather use a probabilistic acceptance, in which
we accept the perturbed parameter with a probability that depends on
$\epsilon$; if it is not accepted, we keep the present value of the
parameter.  The algorithms belonging to the explicit acceptance group are
RejectionABC and PMCABC, whereas the algorithms in the probabilistic
acceptance group are SMCABC, RSMCABC, APMCABC, SABC and ABCsubsim.  Note
that algorithms with an explicit acceptance step are usually much less
efficient computationally, although they come with more theoretical
guarantees.  In fact, for each iteration you may need to perform the forward
simulations many times, so that there is no way to know in advance how much
time the algorithm will take overall.

In the same way as PMCABC, all sequential sampling schemes exploit a
perturbation kernel to explore the parameter space.  In \pkg{ABCpy}, we
usually refer to this simply as $kernel$; \pkg{ABCpy} implements a
multivariate Normal or multivariate Student's-$T$ for continuous variables,
and a random walk kernel for discrete ones.  It is also possible to specify
different kernel functions for different subsets of the parameters, as
described in Section~\ref{sec:comp_pert_ker}.

We also implement the population Monte Carlo (PMC) \citep{Cappe_2004} and the standard Metropolis-Hastings Markov Chain Monte Carlo \citep{Hastings}
sampling schemes to be used with the different likelihood approximations
discussed in Section~\ref{sec:appr_lik}.  A detailed description of PMC
algorithm is provided in Algorithm~\ref{alg:PMC}.  Note that, similarly to
sequential ABC algorithms, PMC sampling scheme also uses a perturbation
kernel to explore the parameter space (Line~11 in Algorithm~\ref{alg:PMC}).
\begin{algorithm}[t!]
\caption{PMC algorithm using an approximate likelihood function and
  producing $N$ samples from the approximate posterior distribution.  Here
  $K_t(\cdot \mid \theta, \Sigma_{t-1})$ is the perturbation kernel, and
  \code{weigthed-Covariance} (not shown here) updates the covariance matrix
  of the perturbation kernel according to the drawn samples and weights.}
\label{alg:PMC}
\begin{algorithmic}[1]
\REQUIRE Specify $\hat{L}_{\datasim}(\cdot \mid \theta)$ function to evaluate
approximate likelihood function at $\theta$ using simulated data $\datasim$.
\FOR{$i=1$ \TO $N$}
\STATE Generate $\theta$ from the prior $\prior(\cdot)$
\STATE Generate $\datasim$ from $\Model$ using $\theta$
\STATE $\ith\theta_1 \leftarrow \theta$
\STATE $\ith\omega_1 \leftarrow
\prior(\theta)\hat{L}_{\datasim}(\dataObs \mid \theta)$
\ENDFOR
\STATE $\Sigma_1 \leftarrow 2*\mbox{\code{weighted-Covariance}}(\theta_1, \omega_1)$ 
\FOR{$t=2$ \TO $T$}
\FOR{$i=1$ \TO $N$}
\STATE Draw $\theta^*$ from among $\theta_{t-1}$ with probabilities $\omega_{t-1}$
\STATE Generate $\theta$ from $K_t(\theta^*,\Sigma_{t-1})$
\STATE Generate $\datasim$ from $\Model$ using $\theta$
\STATE $\ith\theta_t \leftarrow \theta$
\STATE $\ith\omega_t \leftarrow
\prior(\theta)\hat{L}_{\datasim}(\dataObs \mid \theta)/
(\sum^N_{k=1}\kth\omega_{t-1}K_t(\theta \mid \kth\theta_{t-1},
\Sigma_{t-1}))$
\ENDFOR
\STATE Normalize $\ith\omega_t$ over $i = 1,\ldots, N$
\STATE $\Sigma_t \leftarrow 2*\mbox{\code{weighted-Covariance}}(\theta_t,
\omega_t)$
\ENDFOR
\end{algorithmic}
\end{algorithm}

\section[ABCpy]{\pkg{ABCpy}} \label{sec:ABCpy_sructure}

\begin{figure}
\centering
  \includegraphics[width=0.80\textwidth]{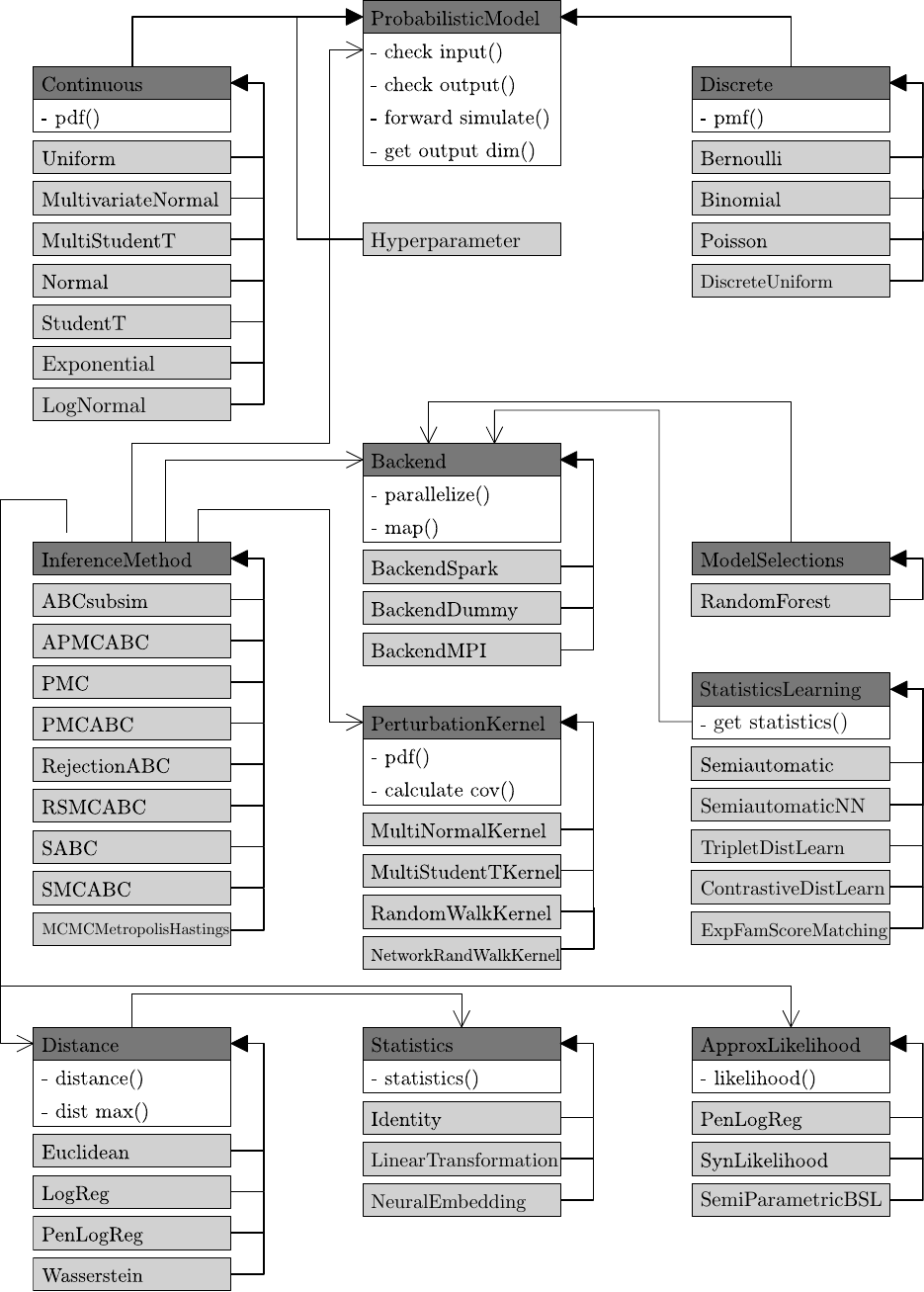}
\caption{This diagram shows selected classes with their most important
  methods.  Abstract classes, which cannot be instantiated, are highlighted
  in dark gray and derived classes are highlighted in light gray. 
  Inheritance is shown by filled arrows.  Arrows with no filling highlight
  associations, e.g.,~\code{Distance} is associated with \code{Statistics}
  because it calls a method of the instantiated class to translate the input
  data to summary statistics.  \code{ContrastiveDistLearn},
  \code{TripletDistLearn}, \code{MultiNormalKernel} and
  \code{MultiStudentTKernel} are used as an abbreviation for
  \code{ContrastiveDistanceLearning}, \code{TripletDistanceLearning},
  \code{MultivariateNormalKernel} and \code{MultivariateStudentTKernel}
  respectively.}
  \label{fig:class-diagram}
\end{figure}

First we give a brief overview of how the \pkg{ABCpy} package works and how
it is used.  Note that \pkg{ABCpy} is under active development and thus the
presented API is prone to changes.  All coded examples work against major
version 0.5.x {and 0.6.x} of \pkg{ABCpy}.  As described in Section
\ref{sec:ABC}, the fundamental components required by ABC methods are:
\begin{itemize}
\item observed data $\dataObs$
\item simulator-based model $\Model$
\item prior distribution $\prior(\theta)$
\item summary statistics
\item discrepancy measure (\emph{distance}) or approximate likelihood
function ($approx\_lhd$)
\end{itemize}
Though not standard for \proglang{Python}, we implemented abstract classes
to define a clear application programming interface (API) on how to use and
extend the library (see Figure~\ref{fig:class-diagram}).  The abstract
classes reflect, among others, the components above:
\begin{itemize}
\item \code{ProbabilisticModel} defines how to provide methods to simulate
data given parameters $\theta$
\item \code{Statistics} defines how to provide methods to extract statistics
\item \code{Distance} defines how to provide distance calculations
\item \code{ApproxLikelihood} defines how to provide a likelihood
approximation
\end{itemize}
All provided components derive from these abstract classes and implement the
required methods; moreover, the user can easily extend the library by
sub-classing the above abstract classes.

In \pkg{ABCpy}, the \class{abcpy.probabilisticmodels.ProbabilisticModel}
class represents the probabilistic relationship between random variables or
between random variables and observed data.  Each of the
\code{ProbabilisticModel} objects has a number of input parameters: they are
either random variables (output of another \code{ProbabilisticModel} object)
or constant values known to the user (of type \code{Hyperparameter}).

To define the parameter of a model as a random variable, the user has to
assign a \emph{prior distribution} on it.  To this aim they can exploit
\emph{prior} knowledge about the parameter value and its distribution.  In
the absence of prior knowledge, we still need to provide prior information
and a flat distribution on the parameter space can be used.  The prior
distribution on the random variables are assigned by a probabilistic model
which can take, as inputs, either other random variables or
hyper-parameters.

We consider now the Lorenz95 model as discussed in
Section~\ref{sec:Introduction}.  Assuming we observe a realization of the
model, we are interested in inferring two one-dimensional parameters ($
\theta_1, \theta_2 $) that enter in the definition of the equations; more
information on the structure of the model is given in
Appendix~\ref{app:lorenz}.  We define the graphical structure of the model
as follows\footnote{The code needed to run this and the following examples
are provided in the supplementary material (this also includes the
definition of the \code{StochLorenz95} model and the
\code{HakkarainenLorenzStatistics} statistics used later in the text, not
shown here for brevity).}:
\begin{CodeChunk}
\begin{CodeInput}
>>> from abcpy.continuousmodels import Uniform
>>> theta1 = Uniform([[0.5], [3.5]], name = "theta1")
>>> theta2 = Uniform([[0], [0.3]], name = "theta2")
>>> sigma_e = 1; phi = 0.4; T = 1024
>>> lorenz = StochLorenz95([theta1, theta2, sigma_e, phi, T],
...   name = "lorenz")
\end{CodeInput}
\end{CodeChunk}
We have thus defined the parameter $\theta_1$ and $\theta_2$ of the Lorenz95
model as random variables and have specified Uniform prior distributions for
them.  The parameters of the prior distribution and the parameters $
\sigma_e $ and $ \phi $ of the model are assumed to be known to the user,
hence they are called hyper-parameters.  Also, internally, the
hyper-parameters are converted to \code{Hyperparameter} objects.  Finally, $
T $ defines the number of integration timestep used for the model.

Note that you can pass a name string (e.g, \code{"theta_1"}) while defining
a random variable.  In the final output, you will see these names, together
with the relevant outputs corresponding to them.

As the output of each integration of the model is a 40 dimensional
timeseries with $ T $ steps, it is computationally inefficient to apply ABC
inference on the output directly.  Therefore, we extract a six-dimensional
set of summary statistics suggested in~\cite{Hakkarainen_2012} before
computing the discrepancy measure as the Euclidean distance between
statistics of different realizations.  The definition of these summary
statistics looks as follows (the class definition is reported in the full
example):
\begin{CodeChunk}
\begin{CodeInput}
>>> statistics_calculator = HakkarainenLorenzStatistics(degree = 1,
...   cross = False)
\end{CodeInput}
\end{CodeChunk}
The discrepancy measure is defined in the next piece of code and takes as
argument the corresponding statistics; when the inference algorithm is run,
it will automatically extract the statistics from the datasets and
subsequently compute the distance between the two statistics.
\begin{CodeChunk}
\begin{CodeInput}
>>> from abcpy.distances import Euclidean
>>> distance_calculator = Euclidean(statistics_calculator)
\end{CodeInput}
\end{CodeChunk}
{As discussed in Section~\ref{sec:ABCpy_alg}, most algorithms in \pkg{ABCpy}
(except for \code{RejectionABC})} require a perturbation kernel to explore
the parameter space.  For this example we use the default kernel, which in
the case of continuous parameters uses a multivariate Gaussian distribution;
it can be defined in the following way:
\begin{CodeChunk}
\begin{CodeInput}
>>> from abcpy.perturbationkernel import DefaultKernel
>>> kernel = DefaultKernel([theta1, theta2])
\end{CodeInput}
\end{CodeChunk}
Finally, we need to specify a backend that determines the parallelization
framework to use.  The example code here uses the \pkg{MPI} backend
\code{BackendMPI} which parallelizes the computation of the inference
schemes using \pkg{MPI}.  As mentioned earlier, a parallelization backend
supporting \pkg{Spark} (\code{BackendSpark}) is available, as well as a
dummy one (\code{BackendDummy}) which does not parallelize the computations,
but is handy for prototyping and testing.  A detailed description of how the
parallelization schemes work is in the Section~\ref{sec:Parallelism}.
\begin{CodeChunk}
\begin{CodeInput}
>>> from abcpy.backends import BackendMPI as Backend
>>> backend = Backend()
\end{CodeInput}
\end{CodeChunk}
For the sake of illustration we choose the PMCABC algorithm as the inference
scheme to draw posterior samples of the parameters.  Therefore, we
instantiate a \code{PMCABC} object by passing the model, the distance
function, backend object, perturbation kernel and a seed for the random
number generator.
\begin{CodeChunk}
\begin{CodeInput}
>>> from abcpy.inferences import PMCABC
>>> sampler = PMCABC([lorenz], [distance_calculator], backend, kernel,
...   seed = 1)
\end{CodeInput}
\end{CodeChunk}
Finally, we can parametrize the sampler by specifying the number of steps
\code{steps}, the number of posterior samples \code{n_samples} and the
number of simulations for each parameter value \code{n_samples_per_param}:
\begin{CodeChunk}
\begin{CodeInput}
>>> steps, n_samples, n_samples_per_param, full_output = 3, 10000, 1, 1
>>> eps_arr = np.array([500]); eps_percentile = 10
\end{CodeInput}
\end{CodeChunk}
Note that the \pkg{ABCpy} implementation of the PMCABC algorithm
(Algorithm~\ref{alg:PMCABC}) is parametrized with an array of threshold values
$(\epsilon_t)_t$ {(\code{eps_arr} here)} and a percentile value
{(\code{eps_percentile})}, and that at iteration $t $ of the algorithm the
actual threshold will be the maximum between $\epsilon_t$ and the percentile
of the distances from the previous iteration (see
Algorithm~\ref{alg:PMCABC}).  \pkg{ABCpy} allows however to specify only the
first threshold values, in which case the iterations starting from the
second one will use the percentile of the previous iteration distances.

We can now sample from the posterior distribution of the parameters given
the observed dataset \code{observation}:
\begin{CodeChunk}
\begin{CodeInput}
>>> journal = sampler.sample([observation], steps, eps_arr, n_samples, 
...   n_samples_per_param, eps_percentile, full_output = full_output)
\end{CodeInput}
\end{CodeChunk}
The above inference scheme gives us samples from the posterior distribution
of the parameters \code{theta_1} and \code{theta_2}, implicitly quantifying
the uncertainty of the inferred parameter, which are stored in the
\code{journal} object.  In particular the posterior mean and covariance
matrix of $(\theta_1, \theta_2)$ are obtained as:
\begin{CodeChunk}
\begin{CodeInput}
>>> print(journal.posterior_mean())
>>> print(journal.posterior_cov())
\end{CodeInput}
\end{CodeChunk}
A plot for the bivariate and univariate marginals posterior distributions
can be obtained and saved to the disk with:
\begin{CodeChunk}
\begin{CodeInput}
>>> journal.plot_posterior_distr(path_to_save =
...   "../Figures/lorenz_hakkarainen_pmcabc.pdf")
\end{CodeInput}
\end{CodeChunk}
Note that the model and the observations are given as a list.  This is due
to the fact that in \pkg{ABCpy}, it is possible to have hierarchical models
and to build relationships between co-occurring groups of datasets, as
detailed in Section.~\ref{sec:hier_mod}.

\section{Modular API} \label{sec:modular}

As one can notice from the structure of the code, the design of \pkg{ABCpy} is highly
modular, so that adapting to different use cases and scenarios can be done with
as little overhead as possible.  In this section, we show how \pkg{ABCpy}'s
modularity addresses the needs of various use cases in a user-friendly,
intuitive way. The contributions to each use case are detailed as follows:
\begin{enumerate}
\item Non-ABC experts do not have to worry about the details of the sampling
  scheme; no knowledge of the interaction between sampling schemes, models,
  kernels etc. is needed.
\item Non-HPC experts can easily run the ABC schemes on hundreds of cores
  even without explicitly parallelizing their code.
\item ABC experts can easily extend the library with new ABC algorithms
  (rapid prototyping) and compare their performance in a standardized
  environment.
\end{enumerate}
Scientists who want to use ABC to calibrate their models only need an
abstract understanding of the ABC methodology and only need to provide
information in the domain of their expertise.  The model and the means to
forward simulate data for given model parameters are the most fundamental
information they need to provide.  Further, scientists usually have a way to
discriminate two simulation outcomes and can make an informed decision on
which better fits the observed data.  This knowledge domain expertise can
drive the choice of the ABC summary statistics.  Apart from this, the user
only has to provide prior information and parametrizations of the sampling
scheme.  These include a perturbation kernel, simulation length and
simulation stopping criteria.  All ABC details are completely handled by the
corresponding modules.

ABC experts can extend the library by providing new sampling schemes,
distances or approximate likelihood methods.  {To do so, the user can
sub-class the \code{InferenceMethod}, \code{Distance} or
\code{ApproxLikelihood} abstract classes and implement the relevant methods. 
Those can be subsequently used with any \code{ProbabilisticModel} and
\code{Backend}, providing simple and fixed environment for benchmarking and
for testing reproducibility.  Moreover, we provide implementations of
several fundamental ABC algorithm (PMCABC, SMCABC, RejectionABC), which can
be used as a starting point to rapidly prototype similar ones.  For
instance, a new SMCABC-type algorithm can be added by adapting the relevant
lines of code in our SMCABC implementation.}

HPC-experts can adapt the library to their specific system.  For example, in
case \pkg{Apache Spark} or \pkg{MPI} is not available or suitable, a system
engineer might extend the library to available parallel architecture by
sub-classing the \code{Backend} class.

\subsection{API design decisions} \label{sec:API_design}

In this section, we provide some background on what led to current design
decisions, in particular why we chose \proglang{Python}, \pkg{MPI}, \pkg{Spark},
and the map-reduce paradigm.

Let us first explain why \proglang{Python} was selected over other
languages.  For high-level scripting languages, \proglang{Python} is {one of
the most used languages} in data science.  It comes with a large range of
well-tested scientific libraries, such as \pkg{NumPy}
\citep{harris2020array} and \pkg{SciPy} \citep{2020SciPy-NMeth}.  Further,
if one considers the standard use case of data scientists, usually rapid
prototyping is required rather than finding a solution and then tweaking it
to work optimally to solve the same problem over and over again.  Thus we
chose it against low-level languages such as \proglang{C++} or
\proglang{Fortran}.  Further, in ABC most computation time is spent
simulating from the model.  In case this might be too inefficient in
\proglang{Python}, it can be implemented in a lower level language as
\proglang{Fortran} or \proglang{C++}, and connected to \proglang{Python}
using e.g.,~\pkg{SWIG} \citep{beazley1996swig}, for which we provide
examples in the documentation\footnote{The documentation can be found at
\href{https://abcpy.readthedocs.io/en/latest/}{https://abcpy.readthedocs.io/en/latest/}.}.

The parallelization backend follows the map-reduce programming model.  An
important argument for map-reduce is its simplicity: there is no need to
explicitly handle communication or worry about thread-safety, deadlocks, or
race-conditions.  The price to pay is that not every problem is easily
expressible in a map-reduce fashion.  However, this is not a constraint for
us since the individual tasks of the ABC sampling schemes are more or less
independent and no sophisticated communication patter is required.  We
consider the map-reduce paradigm to be sufficient for the implemented
methods.  This belief is also supported by thes performance measurements
presented in Section~\ref{sec:Parallelism}.

We have implemented two different parallelization backends for the library,
one based on \pkg{Apache Spark} \citep{ApacheSpark} and the other based on
\pkg{MPI} \citep{message2012mpi} with the idea that they account for most of the
computing infrastructure nowadays available to researchers and data
scientists.  \pkg{Apache Spark} is widely used in industry for large scale
data analytics and many computer infrastructure services at universities
also offer \pkg{Spark} clusters to their researchers.  Even if this is not
an option, there are many commercial \pkg{Spark} providers (for instance
Amazon Web Services), some of which even offer free access to researchers. 
On the other hand, many high performance clusters found at supercomputing
centers use \pkg{MPI} as a communication framework, which is often optimized to
the respective infrastructure.  To enable users of such facilities to easily
adopt and experiment with \pkg{ABCpy}, we also implemented an \pkg{MPI} backend.

\section{Parallelism} \label{sec:Parallelism}

{As discussed in Section~\ref{sec:ABCpy_alg}, the different sampling schemes
implemented in \pkg{ABCpy} follow a similar flow of instructions}.  Thus, to
explain how the parallelism works, we first refer to Algorithm
\ref{alg:PMCABC}.  The flow of the main loop is as follows:
\begin{itemize}
\item[(i)] (re-)sample a set of parameters $\theta$ either from the prior or
  from an already existing set of parameters (Lines~3, 16, code block);
\item[(ii)]\label{item:step_ii} for each parameter, perturb it using the
perturbation kernel, simulate the model and generate 
pseudo-data, 
compute the distance
between generated and observed data, and either accept 
the parameter value
if the distance is
``small'', or repeat the whole second step (Lines~4--7, 17--21, code block); 
\item[(iii)] for each parameter value calculate its corresponding weight (Lines~8, 22, code block); 
\item[(iv)] normalize the weights, calculate a covariance matrix and a quantile (Lines~10, 24--26, code block).
\end{itemize}
These four steps are repeated until the weighted set of parameters,
interpreted as an approximation of the posterior distribution, converges. 
There are several ways to define ``convergence''; however, we will not go into
the details here.  {See Section~\ref{sec:convergence_checks} for \pkg{ABCpy}
tools to assess convergence.}

Parallelization of the algorithms is done in the following way: resampling
the parameters in step (i) and the small computations in step (iv) are
usually quite fast, even for large numbers of parameters, and thus we
refrain from parallelizing them.  On the other hand, step (ii) and (iii) are
the computationally expensive parts.  The generation of simulated data from
the model, for a given parameter value, usually requires substantial
computational resources.  This step therefore has the highest potential for
parallelization.  As already mentioned, we parallelize in a map-reduce
fashion \citep{Dean2008Mapreduce}.  Therefore, we created a mapping function
that maps each parameter value to a perturbed parameter value and next to a
pseudo-observation $\datasim$ generated from the model with the
corresponding perturbed parameter value.  With this, we can create one task
for each parameter such that step (ii) can be fully parallelized.  The
results of the mapping phase, i.e.,~the accepted parameters, are then
collected by (sent back to) the master.  The weight computation in step
(iii) has a quadratic time complexity in the number of parameters.  Thus, we
again parallelize it by mapping the parameters to their weights.

Usually the parallelized steps (model simulation and weight computation)
take sufficient time so the communication overhead plays only a minor role
in the overall execution time.  Further, in both steps, all tasks can be run
independently of each other since they do not require any communication. 
One would thus expect nearly linear scalability, at least as long as the
inherently sequential parts of the program have a run time much shorter than
the parallel parts.

Map-reduce assumes an underlying master / worker architecture, where the
master orchestrates the work, performs light-weight operations, and
distributes independent tasks to a large set of worker nodes; each worker
can usually run tasks in parallel using \emph{executors} (for instance,
different processors).  In a map phase, the master sends a task in form of a
function to the workers, whose executors apply it independently to each
element of data local to the worker node.  In a reduce phase, the master
makes the workers reshuffle the data and apply a reduce function to the
data.  As a matter of fact, we only need a very simple implementation of
reduce, i.e.,~a \emph{collect}, that sends the data back to the master
without applying any function.  As mentioned, this paradigm is simple to
implement but has the disadvantage of being limited in its expression
complexity.  Fortunately the presented algorithms can be parallelized quite
easily, as the parallel parts of the algorithms can mostly run independently
from each other, so that worker-to-worker interaction is not needed.

\pkg{Apache Spark} is a sophisticated implementation of map-reduce. 
Creating a parallelization backend using \pkg{Apache Spark} is rather simple
since we can entirely rely on the built-in functions.  The \pkg{Spark}
backend can be seen as a wrapper that connects the \pkg{ABCpy} internal
map-reduce functions to the \pkg{Apache Spark} ones.

Creating an \pkg{MPI} backend for \pkg{ABCpy} is a completely different story,
since \pkg{MPI} only comes with a set of low-level functions that enable nodes to
exchange information in a one-to-one, one-to-many, and many-to-many fashion
with additional control mechanisms.  The map and reduce functions thus have
to be implemented with these low-level primitives.  \pkg{MPI} does not naturally
provide a master / worker architecture.  Instead, we select one node to act
as the master and rest are treated as worker nodes.  \pkg{MPI} does not directly
deal with nodes as entities but instead provides a rank which can be seen as
a process that has been bound to a certain number of cores.  We thus
implement our executors to run on a rank.  In our implementation of the map
phase, the master splits the work into tasks and assigns them to executors
such that every executor performs roughly the same number of tasks (or
ideally the some amount of work).  The collect phase is more easy to
implement since we only require the data to be sent back to the master
without any shuffling.

\subsection{Performance evaluation} \label{sec:comp_spark_mpi}

Here we present a performance evaluation of the parallelized architecture of
the PMCABC algorithm (Algorithm \ref{alg:PMCABC}) by analyzing the
scalability with the \pkg{Apache Spark} and \pkg{MPI} backends using the Lorenz95
model and the PMCABC algorithm, both of which were described in
Section~\ref{sec:ABCpy_sructure}.

Full details about the model and the algorithmic parameters for the
experiments in this and the following sections are reported in
Appendix~\ref{app:lorenz}.

To test scalability, we ran the same experiment using the \pkg{Spark} and
\pkg{MPI} backends on the CSCS (Centro Svizzero di Calcolo Scientifico) super
computer Piz Daint, where we used multi-core nodes each having two Intel
Broadwell processors with 36 cores in total and 64GB RAM each.  We kept the
size of the problem fixed and we scaled up the number of worker nodes from 2
to 32 in powers of 2, leading to experiments being run on 72, 144, 288, 576
and 1152 cores respectively.  We also ran a similar experiment using
\pkg{Spark} on AWS in order to investigate the performance of the library on
a commercial cloud computing platform.  We used ``c4.8xlarge'' instances which
provide an equivalent 36 vCPUs and 60GB RAM each.  Due to the multi-core
architecture of Daint and AWS, the cores here are equivalent to the
executors discussed above.  Further, for the \pkg{MPI} backend to be comparable to
\pkg{Spark}, we did not perform any computation on the cores belonging to
the first node and dedicated it to be a Master node.

To study scalability, we considered two quantities: speedup and efficiency. 
The \emph{speedup} $\speedup{\mathcal{A}}{n}$ of a parallel algorithm
$\mathcal{A}$ on $n$ cores with respect to a baseline (number of cores) $m,
m \le n$, is the ratio of the algorithm's running time $t(m)$ on $m$ cores
and the running time $t(n)$ on $n$ cores, $\speedup{\mathcal{A}}{n} = t(n) /
t(m)$.  The \emph{efficiency} $\eff{\mathcal{A}}{n}$ of an algorithm
$\mathcal{A}$ on $n$ cores is defined as the speedup normalized by the ratio
of $n$ to the baseline $m$, i.e.,~$\eff{\mathcal{A}}{n} =
\speedup{\mathcal{A}}{n} m/ n$.
\begin{figure}[t!]
  \centering
  \subfloat[Performance: speedup]{
  \includegraphics[width=0.45\textwidth]{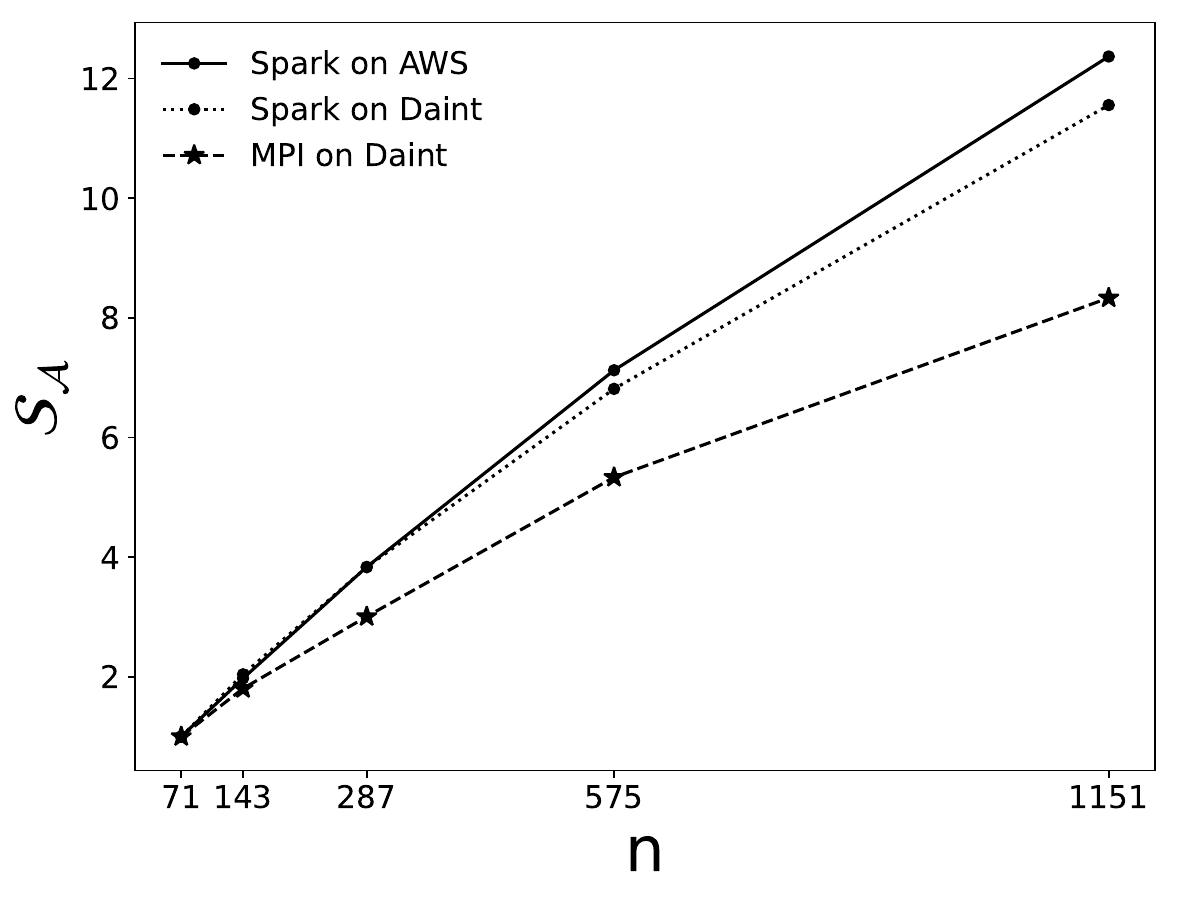}\label{fig:perf_speedup_spark_mpi}}
  \hfill
  \subfloat[Performance: efficiency]{
  \includegraphics[width=0.45\textwidth]{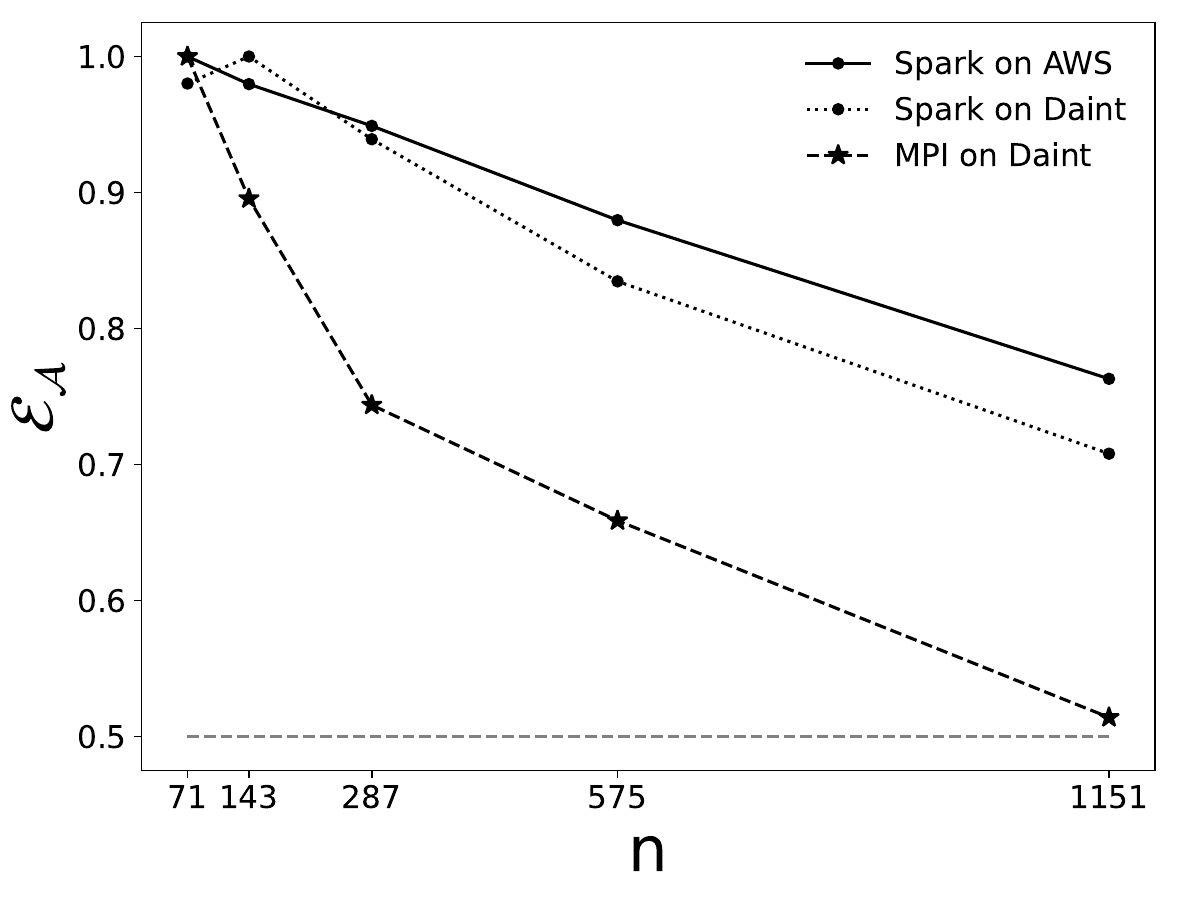}\label{fig:perf_efficiency_spark_mpi}}
   \hfill
  \caption{Speedup $\speedup{\mathcal{A}}{n}$ and efficiency
    $\eff{\mathcal{A}}{n}$ of PMCABC algorithm for Lorenz95 model using
    \pkg{Spark} and \pkg{MPI} backend with different number of cores $n$.  }
  \label{fig:comparison_spark_mpi}
\end{figure}

Figure \ref{fig:comparison_spark_mpi} shows that with the \pkg{Spark}
backend on both Piz Daint and AWS perform similarly.  The performance
increases close to linearly for smaller number of cores but fails to do so
for larger ones.  We attribute this to the fact that the entire process is
not perfectly parallelizable but has serial and parallel regions interlaced. 
As the parallel execution gets faster, the time spent in serial execution
begins to affect overall performance.  Confirming Amdahl's law
\citep{amdahl1967validity}, with increasing parallelism the efficiency
depicted in Figure~\ref{fig:perf_efficiency_spark_mpi} drops as the number
of cores increases.  One can observe that the \pkg{MPI} backend is roughly on par
with the \pkg{Apache Spark} backend in terms of performance, at least up to
576 cores i.e.,~when Amdahl's law starts kicking in.

\subsection[Dynamic allocation for MPI]{Dynamic allocation for \pkg{MPI}} \label{sec:dynamicMPI}

In this Section, we discuss the inherent imbalances of some ABC
algorithms and consequently the importance to study the respective effects. 
As a solution to the imbalance issues, we also discuss the importance of
a dynamic work allocation strategy for map-reduce.  We provide an empirical
comparison of a straightforward allocation approach versus an online greedy
approach.

\begin{figure}[t!]
  \centering
  \subfloat[MPI backend]{
  \includegraphics[width=0.45\textwidth]{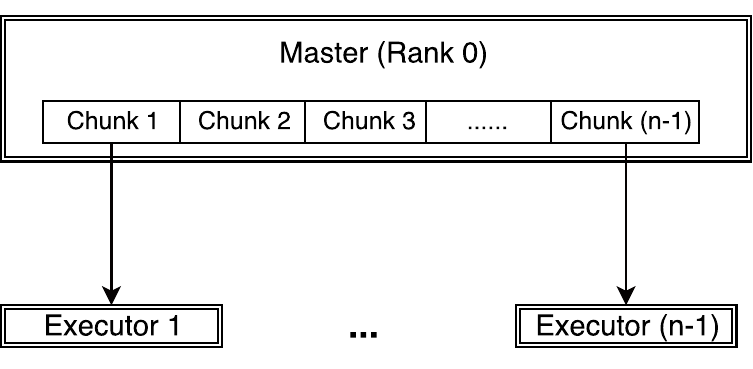}\label{fig:MPI_workflow_basic}}
  \hfill
  \subfloat[dynamic-MPI backend]{
  \includegraphics[width=0.45\textwidth]{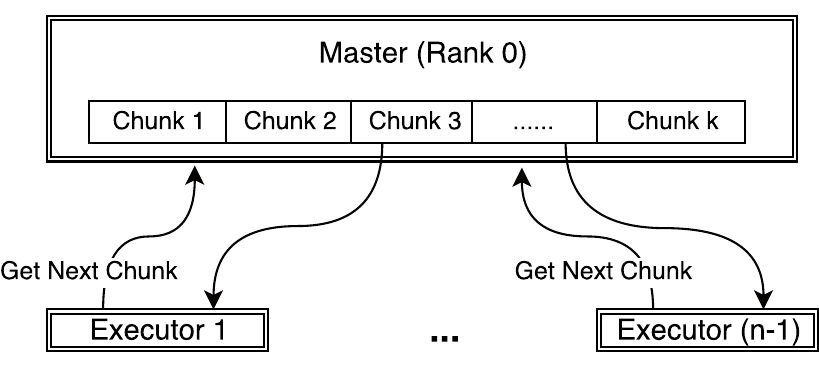}\label{fig:MPI_workflow_dynamic}}
   \hfill
 \caption{Comparison of workflow between \pkg{MPI} and dynamic-\pkg{MPI} backend.}
 \label{fig:MPI_workflow_comparison}
\end{figure}

In the straightforward approach, the allocation scheme initially distributes
$m$ tasks to $n$ executors splitting them identically, and then sends the
map function to each executor, which in turn applies the map function one
after the other for its $m/n$ map tasks.  This approach is visualized in
Figure~\ref{fig:MPI_workflow_basic}, where a chunk represents the set of
$m/n$ map tasks.  For example, if we want to draw $20,000$ samples from the
posterior distribution and we have $n=100$ cores available, at each step of
PMCABC we create chunks of $200$ parameters and each chunk is assigned to
one individual executor.

\begin{figure}[t!]
  \centering
  \subfloat[Performance: speedup]{\includegraphics[width=0.45\textwidth]{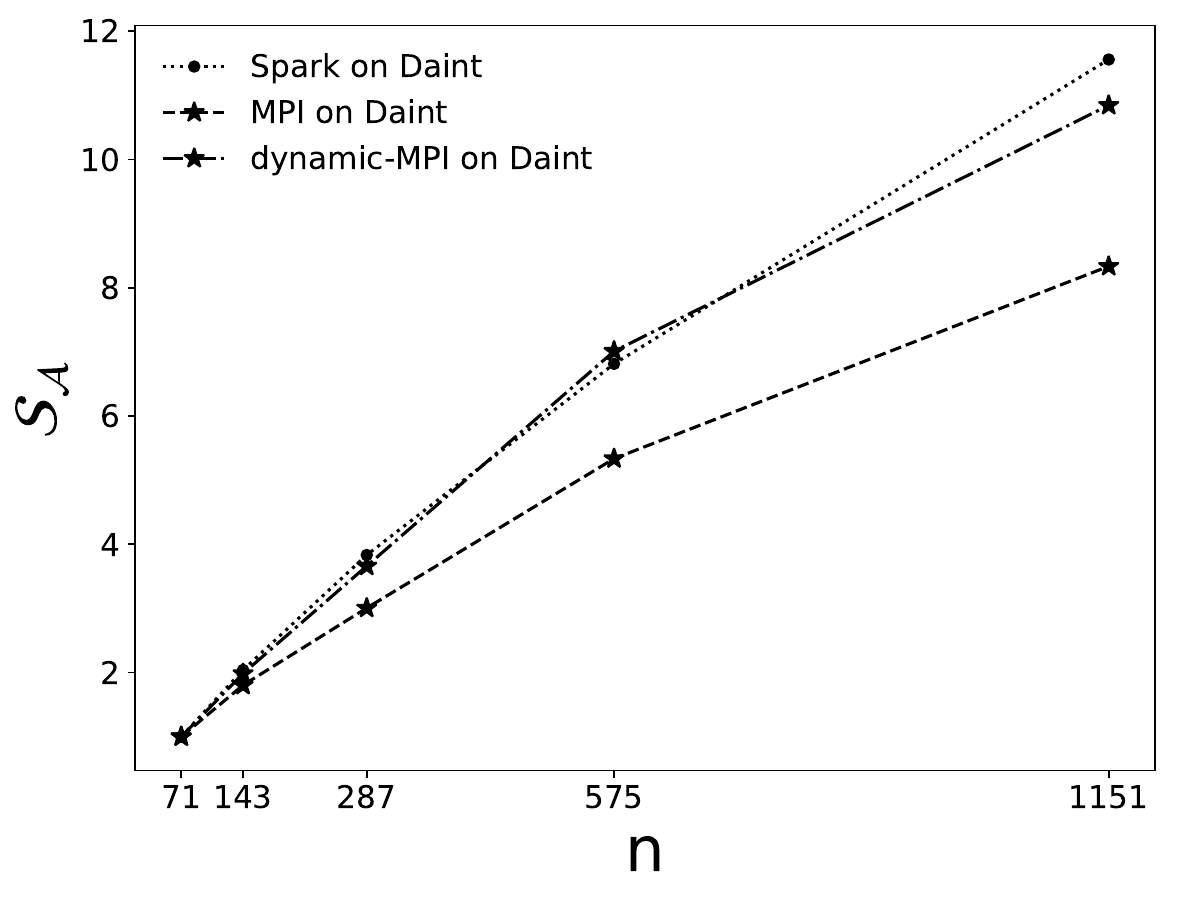}\label{fig:perf_speedup_spark_mpi_dyn}}
  \hfill
  \subfloat[Performance: efficiency]{\includegraphics[width=0.45\textwidth]{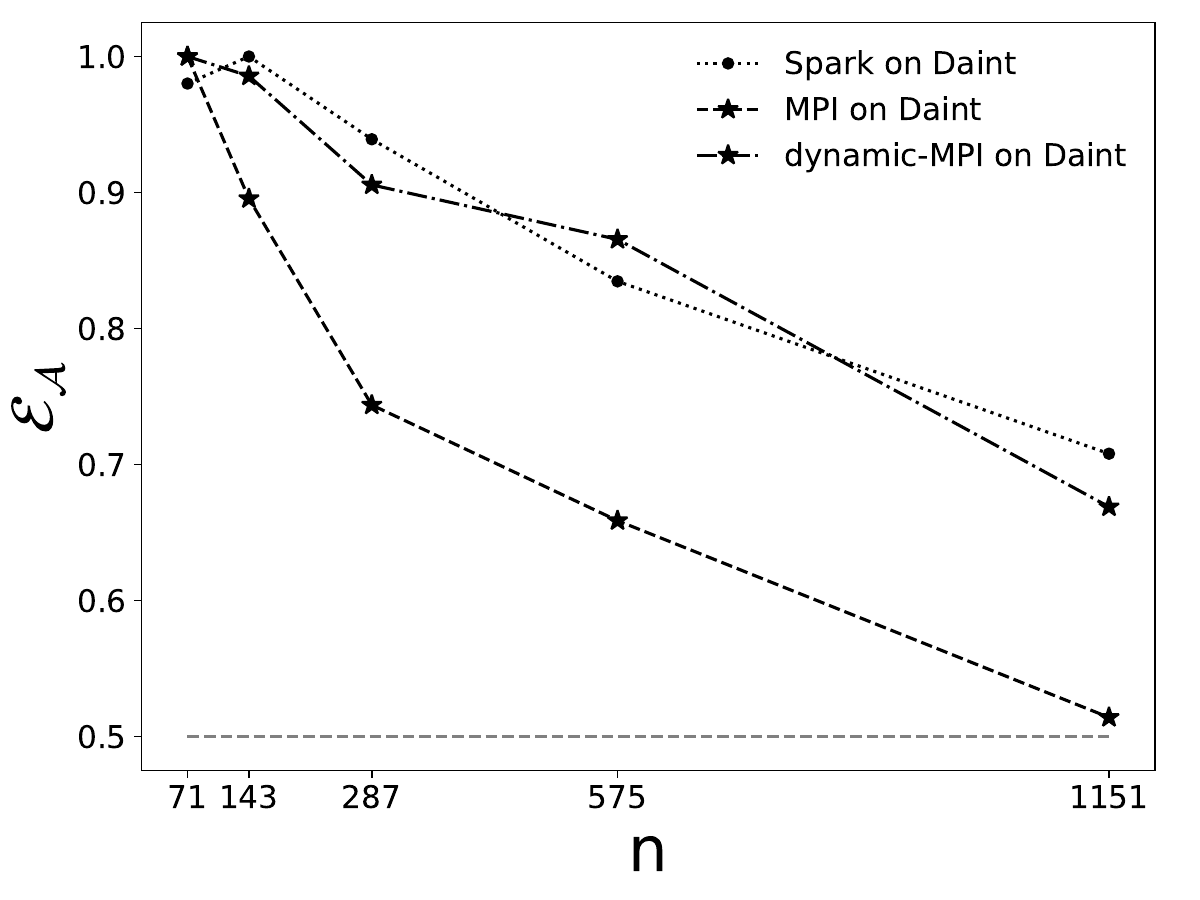}\label{fig:perf_efficiency_spark_mpi_dyn}}
   \hfill
  \caption{Speedup $\speedup{\mathcal{A}}{n}$ and efficiency
  $\eff{\mathcal{A}}{n}$ of PMCABC algorithm for the Lorenz95 model (with $
  T=1024 $) using \pkg{Spark}, \pkg{MPI}(straight-forward) and
  \pkg{MPI}(dynamic-allocation) backends on different number of cores, $n$.}
 \label{fig:comparison_mpi_dyn}
\end{figure}

On the other hand, the dynamic allocation scheme initially distributes $k<m$
tasks to the $k$ executors, sends the map function to each executors, which
in turn applies it to the single task available.  In contrast to the
straightforward allocation, the executor requests a new map task as soon as
the old one is finished.  This has the benefit that the work is better
balanced, as we show in Figure~\ref{fig:comparison_mpi_dyn}.  The dynamic
allocation strategy is an implementation of a greedy algorithm for job-shop
scheduling, which can be shown to have an overall processing time (makespan)
up to twice the best makespan \citep{Graham1966}.  This approach is depicted
in Figure~\ref{fig:MPI_workflow_dynamic}.

\begin{figure}[t!]
  \centering
    \subfloat[MPI(straight-forward)]{
    \includegraphics[width=0.45\textwidth]{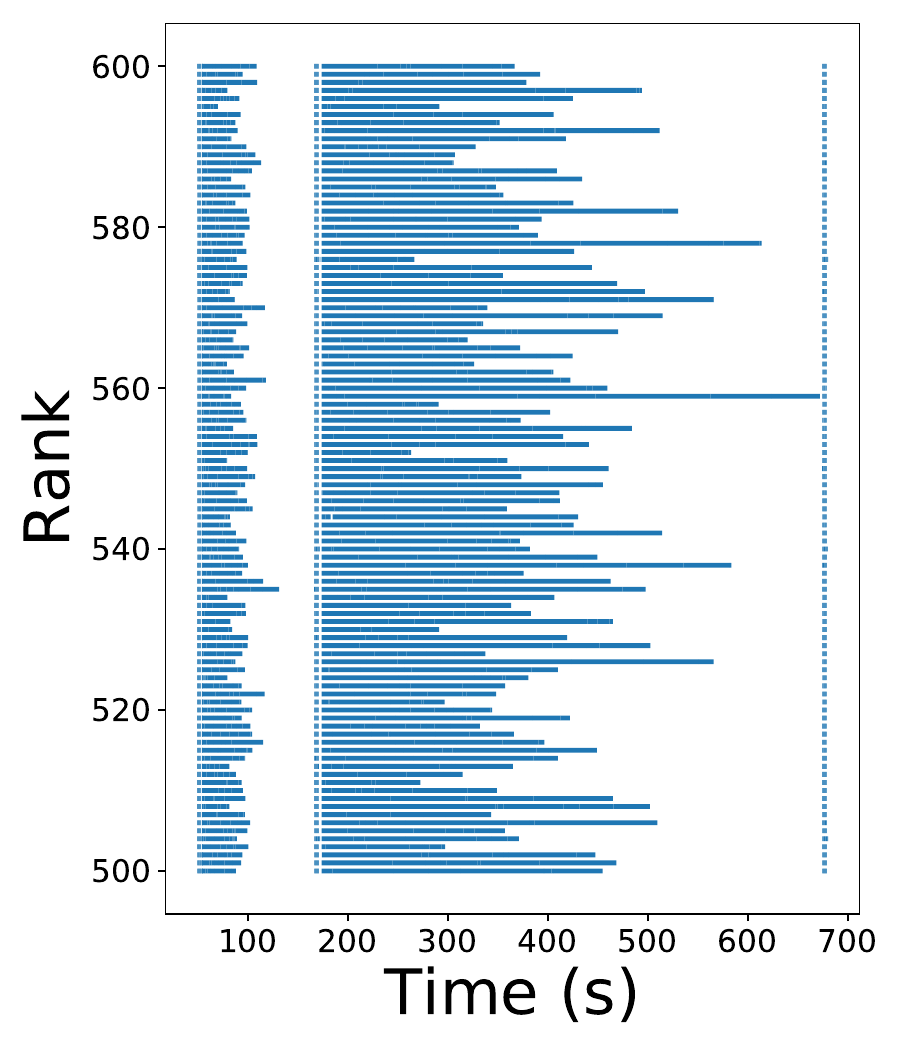}\label{fig:imbalance_pmcabc_mpi}}
    \hfill
    \subfloat[MPI(dynamic-allocation)]{
    \includegraphics[width=0.45\textwidth]{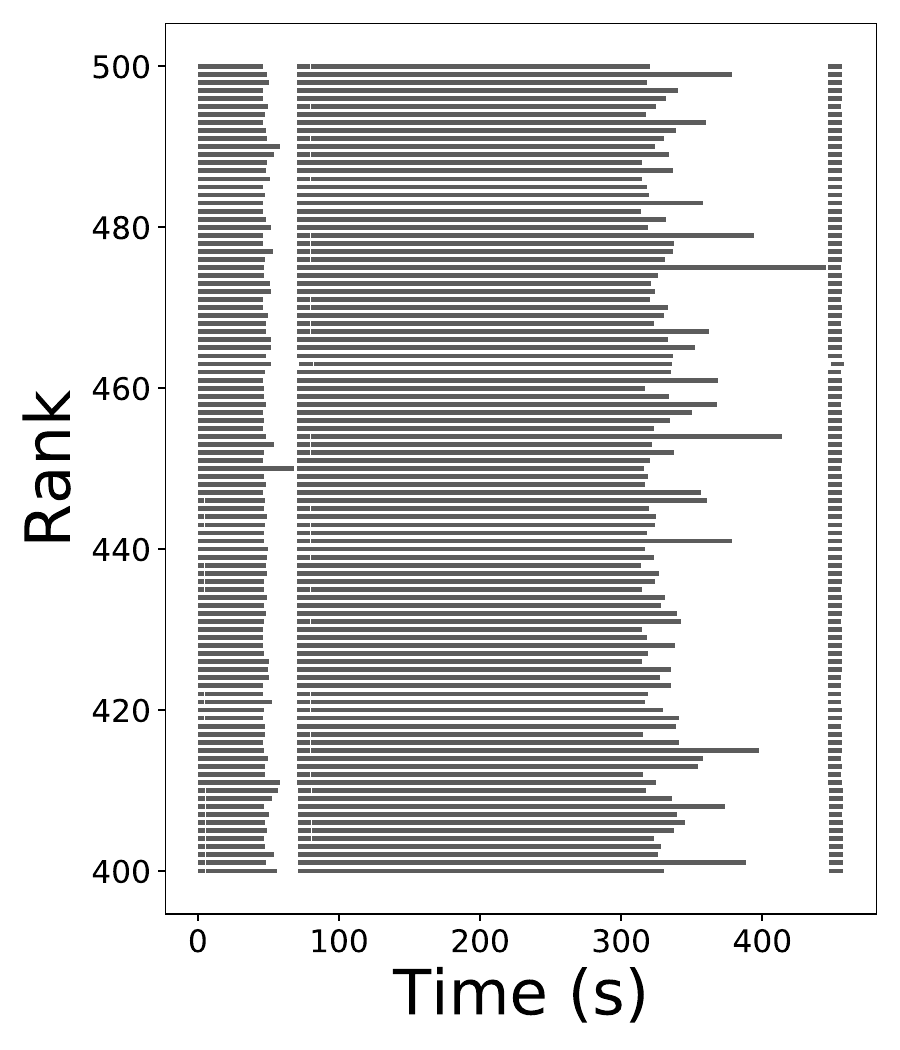}\label{fig:imbalance_pmcabc_dyn}}
  \caption{Imbalance of the PMCABC algorithm using \pkg{MPI}(straight-forward) and
  \pkg{MPI}(dynamic-allocation) backend for the Lorenz95 model ($T=1024$).  Note the
  large difference in the time-scale (in seconds) on the horizontal axis.}
  \label{fig:imbalance_pmcabc}
\end{figure}

The unbalanced behavior can be made apparent by visualizing the run time of
the individual map tasks on each executor.  In Figure~\ref{fig:imbalance_pmcabc},
the individual map task's processing time is
shown for PMCABC.  Each row corresponds to an executor and each bar
corresponds to the total time spent on all tasks assigned to the respective
executor for one map call.  For the straightforward allocation strategy,
Figure~\ref{fig:MPI_workflow_basic}, one can easily see that a majority of
executors finish their map tasks in half the time of the slowest one. 
However, to continue with the next step of the map reduce execution, all
workers and its executors have to be finished.  This clearly leads to large
inefficiencies.  Conversely, using the dynamic allocation strategy,
Figure~\ref{fig:MPI_workflow_dynamic}, the work is more evenly distributed across
the executors.  The cause of the different execution times lie in the
stochasticity of the forward simulation and to a major extent is particular
to the PMCABC algorithm as we discuss later in Section
\ref{sec:imbalance_classify_abc}.

From this observation it follows that the unbalancedness cannot be fixed by
adding resources, and has a severe impact on scalability, as Figure
\ref{fig:comparison_mpi_dyn} shows.  Speed-up and efficiency drop
drastically compared to the \pkg{Spark} implementation and the dynamic
allocation strategy with increasing number of executors.  This can be
understood as follows: in the strong scaling setting, the total number of
map tasks $m$ is fixed, so if we increase the number of executors $k$, the
number of tasks per executor $m/k$ gets smaller.  A small number of map
tasks per executor has a higher variance in the total execution time.

\subsection{Parallelism and ABC algorithms} \label{sec:imbalance_classify_abc} 

In Section~\ref{sec:dynamicMPI}, we pointed out the presence of an inherent
imbalance of the PMCABC algorithm as the execution time of step (ii) for
different parameters varied significantly.  In this section, we explain the
fundamental reason behind this imbalance and then compare different
algorithms in \pkg{ABCpy} from a parallelization perspective.

The acceptance in step (ii) (at Page~\pageref{item:step_ii}) can be easily
split into independent jobs and parallelized for all the algorithms in each
group.  {Recall now the distinction between ABC algorithms with explicit and
implicit acceptance (Section~\ref{sec:ABCpy_alg}); for the latter, one single
simulation for each perturbed parameter value is generated and the parameter
is accepted probabilistically.  In the former case, instead, simulations are
run until the simulation matches the observation at some tolerance level
$\epsilon_t$.} For an explicit acceptance to occur, therefore, it may take
different amounts of time for different perturbed parameters (more repeated
steps are needed if the proposed parameter value is distant from the true
parameter value).  Hence the first group of algorithms are inherently
imbalanced as illustrated for the PMCABC algorithm in
Figure~\ref{fig:imbalance_pmcabc}.  Instead, the algorithms with
probabilistic acceptance do not have a similar issue of imbalance as a
probabilistic acceptance step takes approximately the same amount of time
for each parameter.
 \begin{figure}[t!]
   \centering
   \subfloat[Performance: speedup]{\includegraphics[width=0.45\textwidth]{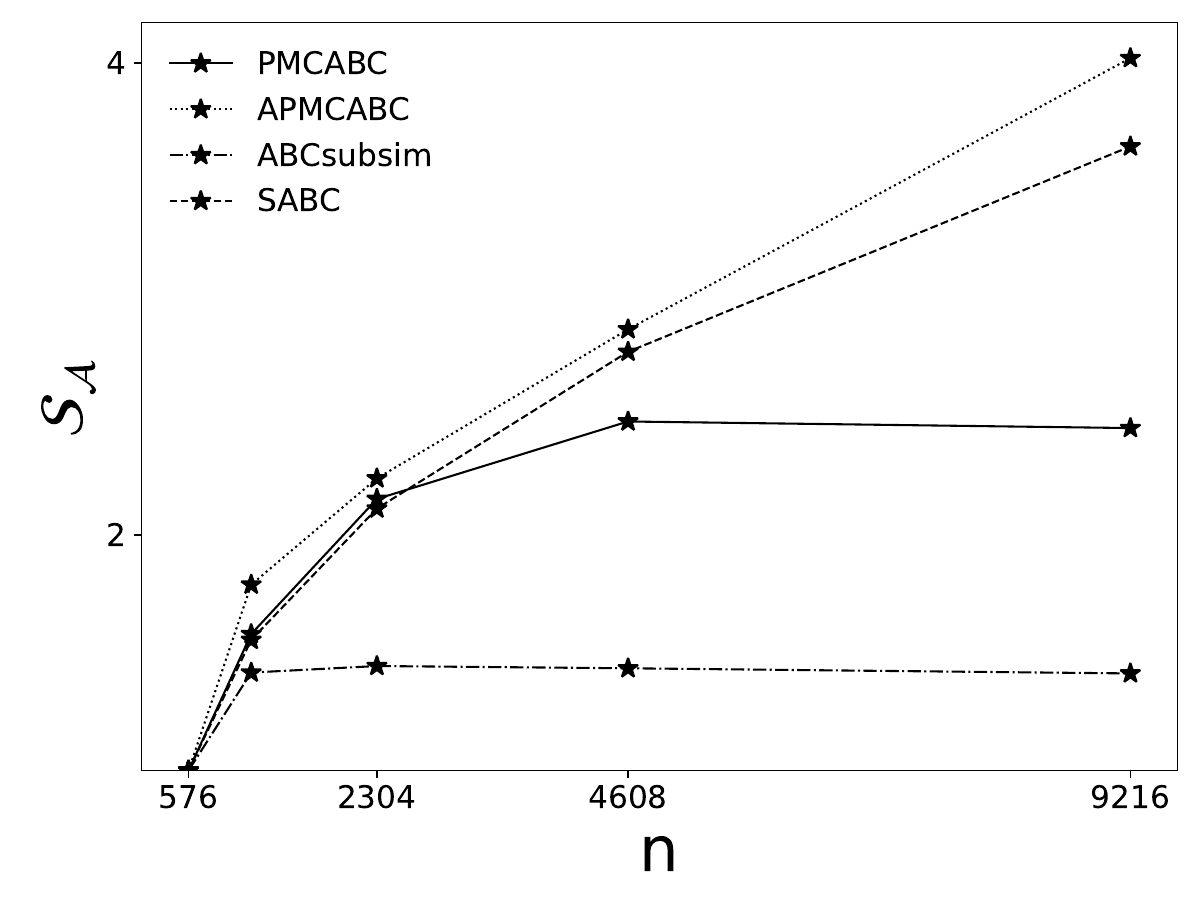}
   }
   \hfill
   \subfloat[Performance: efficiency]{\includegraphics[width=0.45\textwidth]{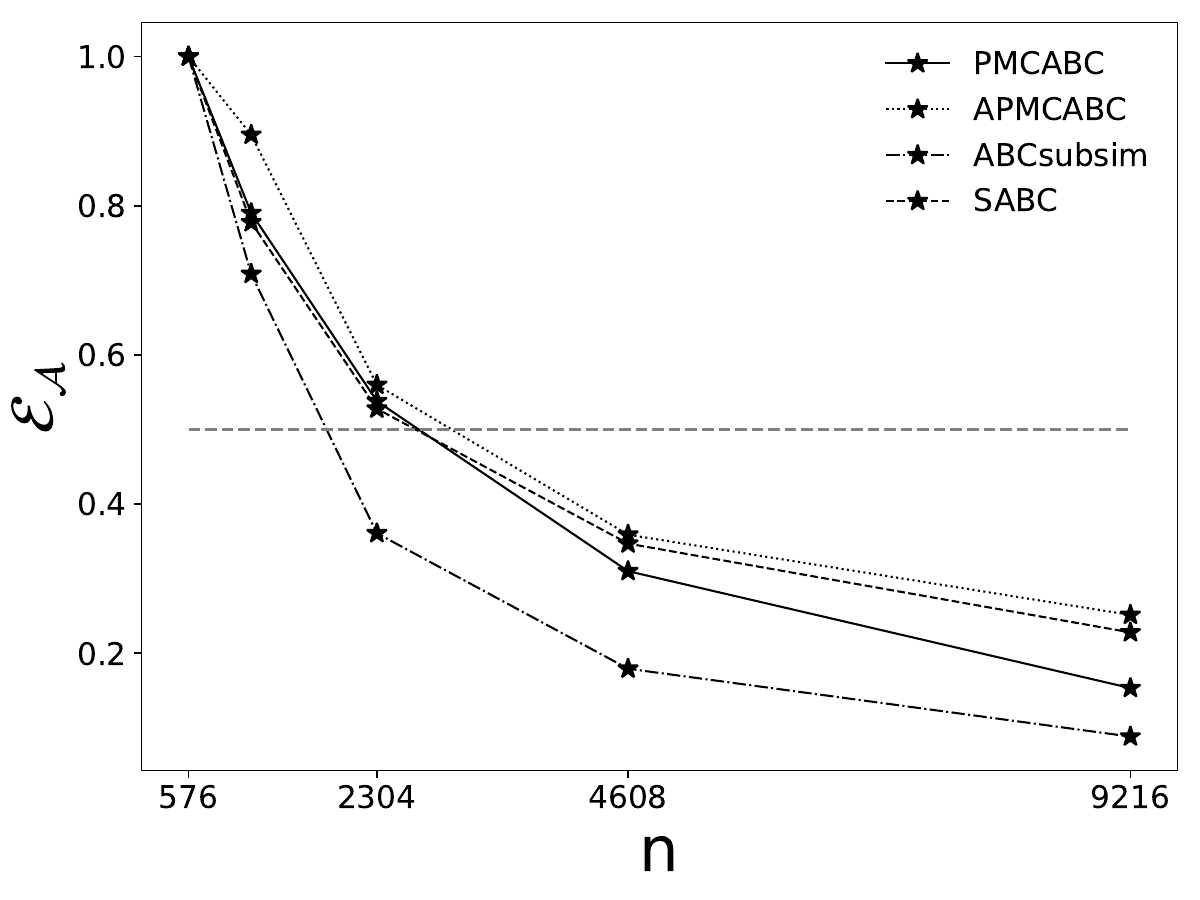}
   }
   \caption{Comparison of speedup and efficiency for PMCABC, SABC, APMCABC
   and ABCsubsim using the Lorenz95 model ($T=1024$).}
   \label{fig:imbalance_pmcabc_sabc_speed_eff}
\end{figure}

Next we compare the achieved performance gain by exploiting parallelism for
four ABC algorithms: PMCABC, APMCABC, SABC and ABCsubsim.  The choice of
these four algorithms were motivated by three aspects: a) PMCABC is the most
classical ABC algorithm; b) APMCABC and SABC are, to the best of our
knowledge, the ABC algorithms with faster convergence to posterior
distribution and the minimal number of model simulations needed
\citep{lenormand2013adaptive, Albert_2015}; c) ABCsubsim is instead a
popular algorithm for engineering applications \citep{Kulakova_2016}. 
{Further we comment that we exclude SMCABC, RSMCABC and RejectionABC from
our analysis, due to the almost similar performance of SMCABC in comparison
to SMCABC and the inability of RSMCABC and RejectionABC to scale up while
using correspondingly more than 1 or 100~parallel cores, which is a much
smaller number with respect to the one we considered here.}

We run now the above algorithms on the Lorenz95 model as discussed in
Section~\ref{sec:ABCpy_sructure}; as the code for PCMABC was provided there,
we show here how to run the inference with the other three algorithms; all
parameters, except for the specified ones, are left to their default value:
%
\begin{CodeChunk}
\begin{CodeInput}
>>> from abcpy.inferences import SABC, APMCABC, ABCsubsim

>>> ## Run inference with SABC
>>> sampler = SABC([lorenz], [distance_calculator], backend, kernel,
...   seed = 1)
>>> # Define sampling parameters
>>> steps, n_samples, n_samples_per_param, full_output = 20, 10000, 1, 1
>>> epsilon = 500
>>> # Sample
>>> journal = sampler.sample([observation], steps, epsilon, n_samples,
...   n_samples_per_param, full_output = full_output)

>>> ## Run inference with ABCsubsim
>>> sampler = ABCsubsim([lorenz], [distance_calculator], backend, kernel,
...   seed = 1)
>>> # Define sampling parameters
>>> steps, n_samples, n_samples_per_param, full_output = 20, 10000, 1, 1
>>> # Sample
>>> journal = sampler.sample([observation], steps, n_samples,
...   n_samples_per_param, full_output = full_output)

>>> ## Run inference with APMCABC
>>> sampler = APMCABC([lorenz], [distance_calculator], backend, kernel,
...   seed = 1)
>>> # Define sampling parameters
>>> steps, n_samples, n_samples_per_param, full_output = 20, 10000, 1, 1
>>> acceptance_cutoff = 0.003
>>> # Sample
>>> journal = sampler.sample([observation], steps, n_samples,
...   n_samples_per_param, full_output = full_output)
\end{CodeInput}
\end{CodeChunk}
In Figure~\ref{fig:imbalance_pmcabc_sabc_speed_eff}, we compare the speed-up
and efficiency of the considered algorithms.  More details on the settings
of the different algorithms can be found in Appendix~\ref{app:lorenz}.  We
notice that ABC algorithms with ``probabilistic acceptance'' do not have an
inherent imbalance, but they may not be easily parallelizable due to the
sequential nature of the algorithm, which is illustrated by the poor
performance of ABCsubsim algorithm compared to the others.  We also conclude
that the performance of APMCABC and SABC is significantly better compared to
PMCABC due to the absence of imbalance in them and are therefore better
suited for a parallelization with the map-reduce paradigm.

\begin{figure}
   \centering
\subfloat[PMCABC]{
	\includegraphics[width=0.45\textwidth]{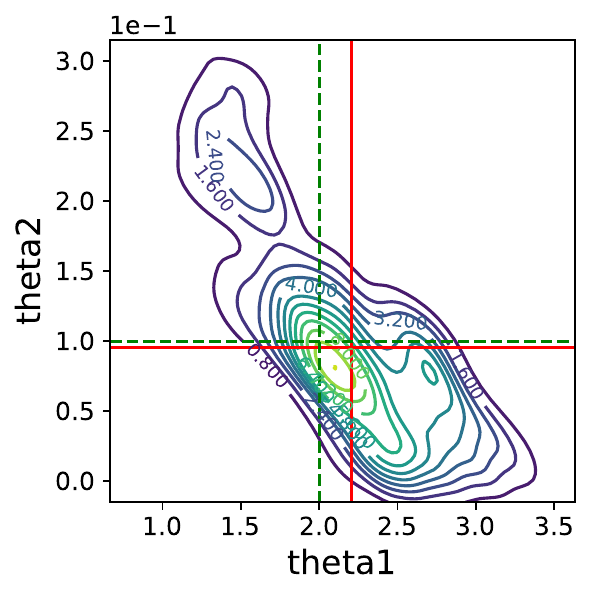}
	\label{fig:Lorenz_posterior_pmcabc}
}
\hfill
\subfloat[APMCABC]{
	\includegraphics[width=0.45\textwidth]{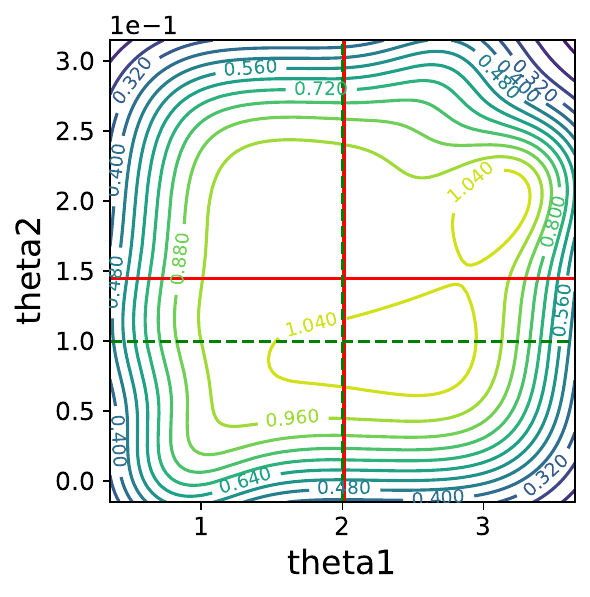}
}
\newline
\subfloat[SABC]{
	\includegraphics[width=0.45\textwidth]{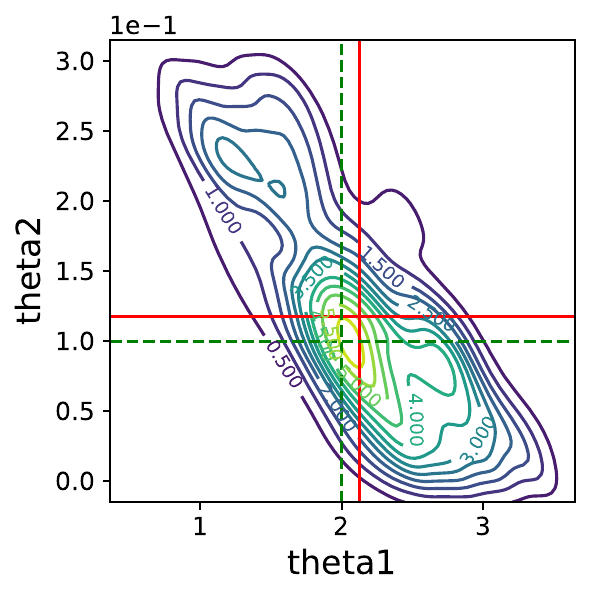}
}
\hfill
\subfloat[ABCsubsim]{
	\includegraphics[width=0.45\textwidth]{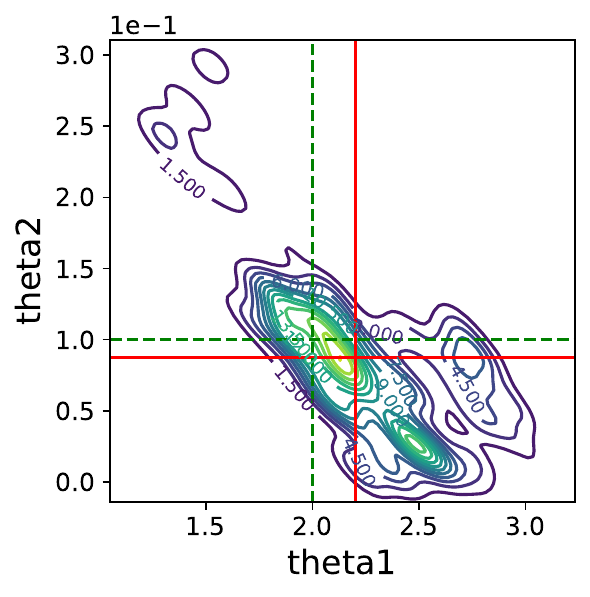}
}
\caption{Contour plots for the approximate posteriors inferred with the
different ABC algorithms.  The red lines denote position of the posterior
mean, while the green lines denote the position of the true parameter
values.  The posteriors were obtained with kernel density estimate starting
from the posterior samples.}
\label{fig:Lorenz_posterior}
\end{figure}

Moreover, with regards to the total computational complexity of the
different algorithms, note that running one of the algorithms implemented in
the above code chunk (not involving explicit acceptance step) for 20
iterations with 10000 posterior sample points took roughly as long as
running the PMCABC for 3 iterations, with the same number of samples (see
code in Section~\ref{sec:ABCpy_sructure}).  In fact, once PMCABC reached the
4th iteration, for each accepted simulation of the model, around 1000
simulations were needed; this is extremely expensive, so that we were not
able to run the algorithm for more than three iterations due to limitations
in computing capability.  This also explains the worse approximation to the
posterior density obtained with this algorithm with respect to the other
ones (Figure~\ref{fig:Lorenz_posterior}).

\section[Innovations of ABCpy compared with similar packages]{Innovations of
\pkg{ABCpy} compared with similar packages} \label{sec:comparison_other_packages}

\begin{table}[t!]\footnotesize
 \centering
\begin{tabular}{@{}l p{0.127\linewidth} p{0.127\linewidth} p{0.127\linewidth}
p{0.127\linewidth} p{0.127\linewidth} @{}}
\hline
\pkg{Feature} & 	\pkg{ABCpy} & \pkg{ELFI} & \pkg{pyABC} & \pkg{abc} &
\pkg{EasyABC} \\
\hline
Language  & \proglang{Python} & \proglang{Python} &  \proglang{Python} &
\proglang{R} & \proglang{R}\\
Latest release	 & 2021  & 2021 & 2021 & 2015 & 2015 \\
Version	 & 0.6.3 & 0.8.0 & 0.11.2 & 2.1 & 1.5 \\
Parallelization	 & Multicore, distributed& Multicore, distributed &
Multicore, distributed & No & Multicore \\
Graph based & Yes & Yes & No & No & No \\
Inference schemes & Wide choice of techniques & BOLFI, SMCABC, RejectionABC
& SMCABC only & RejectionABC only &  Wide choice of techniques \\
Co-occurring dataset & Yes & No & No & No & No \\
Nested parallelization & Yes & No & No & No & No \\
Composite kernel & Yes & No & No & No & No \\
Statistics learning & Yes & No & Preliminary & No & No \\
{Convergence diagnostics }& Yes & Yes & No & No & No \\
\hline
\end{tabular}
\caption{Review of the main features of different ABC packages.}
\label{Tab:comparison}
\end{table}
We now compare \pkg{ABCpy} with other general-purpose ABC packages for
high-level languages, namely \pkg{ELFI} \citep{lintusaari2018elfi} and
\pkg{pyABC} \citep{klinger2018pyabc} for \proglang{Python} and \pkg{abc}
\citep{csillery2012abc} and \pkg{EasyABC} \citep{jaboteasyabc} for
\proglang{R} \citep{R}.  In the following Sections, we highlight the important
innovations that are included in our package and are not available in any of
the competing ones.

In terms of inference techniques, \pkg{ABCpy} is arguably the most complete
one.  In fact, it implements a selection of Sequential and MCMC based
methods, as well as Simulated Annealing ABC.  \pkg{EasyABC} provides a
similar selection, but the latest release was in 2015, therefore missing out
the latest algorithmic developments.  \pkg{ELFI} implements SMCABC,
RejectionABC and BOLFI \citep{gutmann2016bayesian}, that uses Gaussian
process Bayesian optimization to speed up computation.  \pkg{pyABC} only
consider sequential techniques, while \pkg{abc} only provides the
RejectionABC scheme, complemented with two post-processing techniques
\citep{Beaumont2002, blum2010non}.  Moreover, \pkg{ELFI} and \pkg{EasyABC}
are not able to perform model selection.

As discussed above, \pkg{ABCpy} can use \pkg{Spark} and \pkg{MPI} to parallelize
the computation on multicore and distributed systems; the same is possible
with \pkg{ELFI} and \pkg{pyABC}, the former using \pkg{ipyparallel}
\citep{ipyparallel} for distributed systems and \proglang{Python} built in
library for multiple cores, while the latter is able to work with several
backends, among which \pkg{Dask} \citep{dask}, the \pkg{IPython}
\citep{ipython} parallel cluster and \pkg{Redis} \citep{redis}.  We remark
moreover that \pkg{ABCpy} is the only package to offer the nested
parallelization feature, which is detailed in Section~\ref{sec:nested_par}. 
\pkg{ELFI} is moreover able to vectorize simple operations in the simulator,
by performing batches of simulations at once.  \pkg{abc} does not provide
any parallelization, as it assumes the model simulations had been run
beforehand and the output formatted and passed to the package; instead,
\pkg{EasyABC} is able to parallelize only on multicore machines, but if the
simulator code is a binary executable, parallelization requires modifying
it.

The description of the dependencies between the different components of the
probabilistic model, as done in Section~\ref{sec:ABC}, creates an underlying
computational graph in \pkg{ABCpy}.  This allows great flexibility in
specifying an overall model, as different components may be composed in
several ways with no need to changing their structure.  This approach is
also present in \pkg{ELFI}, while it is missing in the other packages
considered here.

With regards to code modularity, we believe \pkg{ABCpy} to be the best
package, alongside with \pkg{ELFI}.  With the exception of \pkg{abc}, that
requires the observation and simulations to be provided to the inference
scheme as matrices and does not allow the implementation of other methods,
the other packages all have a modular structure, but in different ways. 
\pkg{EasyABC} allows models to be specified in functions or external binary
files, but does not separate the model and the statistics component. 
\pkg{pyABC} allows the models to be either functions or classes, but they
need to work with \proglang{Python} dictionaries as input and output. 
Moreover, it is not possible to easily extend \pkg{pyABC} to other inference
schemes, but only to modify the parameters or the scheduling of the SMCABC
algorithm.  \pkg{ELFI} and \pkg{ABCpy} are instead similar in terms of their
modularity.

Additionally, \pkg{ABCpy} implements semiautomatic summary selection
routines, as well as the possibility of using neural networks to learn and
implement statistics in ABC inference; this is described in detail in
Section~\ref{sec:stat_learning}.  The only other package allowing to
automatically learn summary statistics is \pkg{pyABC} (which at the current
release only implements a preliminary version of the semiautomatic method,
with no neural network support).

Also, to the best of our knowledge, \pkg{ABCpy} is the only package
offering the possibility of performing inference with co-occurring
measurements of different quantities that belong to the same graphical model
(see Section~\ref{sec:hier_mod}).  Finally, \pkg{ABCpy} and \pkg{ELFI} are
the only packages to provide tools for assessing the convergence of the
posterior approximation for ABC algorithms
(Section~\ref{sec:convergence_checks}).

In Table~\ref{Tab:comparison} we display a quick summary of the features of
the different packages discussed here.

\subsection{Learning summary statistics} \label{sec:stat_learning}
	
As discussed above, informative summary statistics are a main component of
ABC algorithms.  Practitioners may choose knowledge domain driven summaries,
thus focusing the inferential process on specific data features encoded by
those summaries.  However, in many cases we would like the approximate
posterior to be as close as possible to the one obtained with the whole
dataset, but we still need to use summary statistics as the dimension of the
raw data is too large, leading to poor computational performance.
	
Therefore, ways to automatically learn summary statistics have been
developed.  \pkg{ABCpy} implements some techniques based on mapping the data
to lower dimensional subspaces, that are described in the following.  For
all of them, before the ABC algorithm is run, a set of parameter-simulation
pairs $ (\theta_i, \data_i)_{i=1}^n $ is generated according to the prior
and the model; then, a learning algorithm is applied in order to learn a
data transformation.  During the subsequent inference, the data will be
transformed with the latter, providing the summary statistics.  We note that
before the learning step, the generated data is optionally transformed with
a fixed \emph{statistics} function, for instance to obtain a polynomial
expansion of the raw data.
	
A very popular approach is the one introduced in
\cite{fearnhead2012constructing}, in which the learned transformation is a
linear projection to the dimension of the parameter.  Specifically, the
following linear model is fit:
\begin{equation}\label{}
	\theta_i = \E(\theta_i \mid \data_i) + \xi =  \data_i^\top\beta + \xi,
\end{equation}
where $ \xi $ is a 0-mean noise vector with independent components and
$\beta$ is the set of parameters that are fitted.  During inference,
therefore, statistic for a new sample $\datasim$ will be $ \datasim^\top
\beta $.  This is implemented in the \class{Semiautomatic} class and showed
in the following piece of code:
%
\begin{CodeChunk}
\begin{CodeInput}
>>> from abcpy.statistics_learning import Semiautomatic
>>> from abcpy.statistics import Identity

>>> # summary statistics applied before learning the transformation
>>> statistics_calculator = Identity(degree = 2, cross = True)
>>> # learn now the new summary statistics
>>> new_statistics = Semiautomatic([model], statistics_calculator,
...   backend, n_samples = 200).get_statistics()
\end{CodeInput}
\end{CodeChunk}
Here, note that the Identity statistics applies a polynomial expansion of
order \code{degree} to the data and optionally computes cross products
(argument \code{cross}), before applying the statistics learning algorithm.
	
\begin{figure}[t!]
	\centering
	\subfloat[SABC with hand-defined statistics.]{
		\includegraphics[width=0.45\textwidth]{lorenz_hakkarainen_sabc.pdf}
	}
	\hfill
	\subfloat[SABC with learned statistics.]{
		\includegraphics[width=0.45\textwidth]{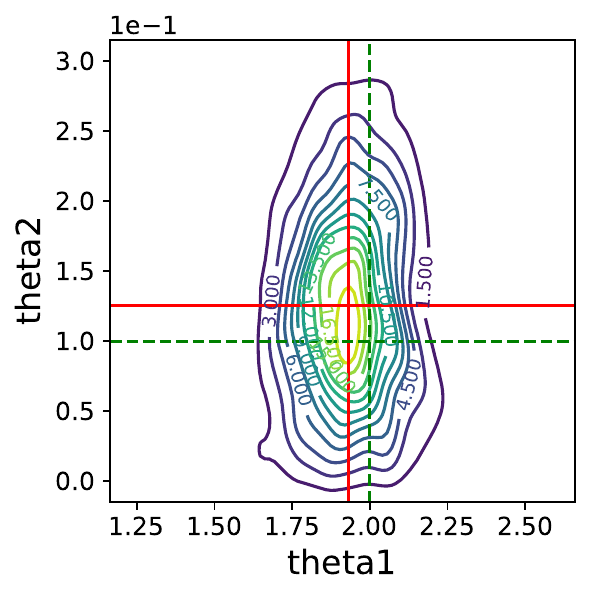}
	} 
\caption{Contour plots for the approximate posteriors obtained with the SABC
algorithm with the hand-chosen statistics defined in
\protect{\cite{Hakkarainen_2012}} and the automatically learned ones.  The
red lines denote position of the posterior mean, while the green lines
denote the position of the true parameter values.  The posteriors were
obtained with kernel density estimate starting from the posterior samples. 
The algorithm with learned statistics is able to concentrate much more
around the true value of $\theta_1$, but not so much around $ \theta_2 $.}
\label{fig:Lorenz_posterior_stats_learning}
\end{figure}

The authors of \cite{jiang2017learning} extended this approach by using a
neural network model instead of a linear transformation, namely replacing $
\data_i^\top\beta $ by $ f_w(\data_i) $ in the above expression, where $ f_w
$ denotes the transformation applied by a specific neural network with
weights $ w $, which are determined by iteratively minimizing the
corresponding least squared regression loss; this is implemented in the
\class{SemiautomaticNN} class.  In the same way as before, the statistic will
therefore be $ f_w(\datasim) $.  The neural network summary selection allows
much more representation power than the linear transformation one, with very
small or no additional lines of code required with respect to the linear
regression one.  We give here an example of this technique for the Lorenz95
model, by using as a neural network the Partially Exchangeable Network
introduced in \cite{wiqvist2019partially}, that is an embedding of the
40-dimensional time series whose output is invariant to permutations in the
input that are characteristic of the Markovianity of the time series; see
\cite{wiqvist2019partially} for more details on that.  After having learned
the statistics, we carry out inference using the SABC algorithm.  The
following piece of code implements both the statistics learning and the
inference step\footnote{The code containing the definition of the neural
network classes \class{PhiNetwork}, \class{RhoNetwork} and \class{PEN1} is
available in the supplementary material.}:
%
\begin{CodeChunk}
\begin{CodeInput}
>>> from abcpy.statistics import Identity, Statistics
>>> from abcpy.statisticslearning import SemiautomaticNN

>>> ## define the statistic that will be applied before learning
>>> ## transformation
>>> preprocessing_statistics = Identity(degree = 1, cross = False)

>>> ## define the neural net to be used. This is the Partially
>>> ## Exchangeable Network
>>> phi_net = PhiNetwork()
>>> rho_net = RhoNetwork(n_parameters=2)
>>> embedding_net = PEN1(phi_net, rho_net, n_timestep = T)

>>> # Run now the SemiautomaticNN algorithm to learn the statistics
>>> summary_selection = SemiautomaticNN([lorenz], preprocessing_statistics,
...   backend, embedding_net, n_samples = 500)

>>> # get the learned statistic
>>> statistics_calculator = summary_selection.get_statistics()

>>> # Re-define distance
>>> distance_calculator = Euclidean(statistics_calculator)

>>> ## Run inference with SABC
>>> sampler = SABC([lorenz], [distance_calculator], backend, kernel,
...   seed = 1)
>>> # Define sampling parameters
>>> steps, n_samples, n_samples_per_param, full_output = 20, 10000, 1, 1
>>> epsilon = 500
>>> # Sample
>>> journal = sampler.sample([observation], steps, epsilon, n_samples,
...   n_samples_per_param, full_output = full_output)
\end{CodeInput}
\end{CodeChunk}
Note that after having learned the statistics, the subsequent sampling
inference step is coded in the same way as the one with the hand-chosen
statistics.  Figure~\ref{fig:Lorenz_posterior_stats_learning} reports the
approximate posterior obtained with APMCABC by using both the learned and
the Hakkarainen statistics used throughout the text.
Additionally, \pkg{ABCpy} also implements a newly proposed technique
\citep{pacchiardi2020distance}, which finds a neural network transformation
$f_w(\cdot)$ that is able to approximately preserve the distance of
parameter space; specifically, by denoting as $d_E$ the Euclidean distance,
we look for $ f_w $ such that $ d_E(\theta_i, \theta_j) \approx
d_E(f_w(\data_i), f_w(\data_j)) $ for any $ i, j $.  The intuition is that
if the distance between the statistics is representative of the distance of
the corresponding parameters, then ABC inference will perform well.  Two
different techniques to achieve this are implemented in the classes
\class{ContrastiveDistanceLearning} and \class{TripletDistanceLearning},
respectively based on comparing pairs and triplets of simulated data when
learning the transformation; please refer to \cite{pacchiardi2020distance}
for more details.  Finally, \pkg{ABCpy} implements the summary statistics
learning approach presented in \cite{pacchiardi2020score}, in which an
exponential family approximation is fit to the likelihood, with two neural
networks representing respectively the natural parameters and the sufficient
statistics of the exponential family; the latter will therefore represent
the sufficient statistics of the best exponential family approximation to
the model.  This is implemented in the class
\class{ExponentialFamilyScoreMatching}.
	
We note that \pkg{ABCpy} uses \pkg{Pytorch} \citep{paszke2017automatic} to
handle the neural networks and the corresponding computations.  The package
allows the user to specify a neural network by either passing
\code{torch.nn} object or by specifying the width and depth of fully
connected layers as a list of numbers; alternatively, a default one can be
used, whose size is determined from the dimension of the data and of the
parameter.  As neural networks are not a fundamental part of the ABC
pipeline, but only an optional preprocessing tool, \pkg{Pytorch} is not a
required dependency of \pkg{ABCpy}; rather, whenever one of the neural
network based routines is called, the code checks if \pkg{Pytorch} is
available and, if not, asks the user to install it.

\subsection{Probabilistic dependency between random variables}
\label{sec:prob_dep_rv}

Since release 0.5.x of \pkg{ABCpy}, probabilistic dependency structures
between random variables can be implemented.  Behind the scene, \pkg{ABCpy}
will represent this dependency structure as a directed acyclic graph (DAG)
on which inference can be performed, in the spirit of graphical models.  New
random variables can be defined through operations between existing random
variables.  To make this concept more approachable, we now exemplify an
inference problem on a probabilistic dependency structure.

Let us assume students of a school took an exam and each received a grade. 
Grades are stored in the variable \code{grades_obs}.  We believe grades
depend on several variables: historical grades average, the average size of
the classes, as well as the number of teachers at the school.
\begin{figure}
\centering
\begin{tikzpicture}
	\node[element_elip,text width=3cm]  (SB) at (1, 2) {School Budget};
	\node[element_elip,text width=2cm]  (CS) at (0, 0) {Class Size};
	\node[element_elip,text width=2.5cm]  (GWAE) at (-4, 0) {Historical mean grade};
	\node[element_elip,text width=2cm]  (NT) at (4, 0) {Number of teachers};
	\node[element_elip,text width=2.5cm]  (FG) at (0, -2) {Final Grade};
	
	\draw [decoration={markings,mark=at position 1 with
    {\arrow[scale=3,>=stealth]{>}}},postaction={decorate}] (SB.south) to (CS.north);
    	\draw [decoration={markings,mark=at position 1 with
    {\arrow[scale=3,>=stealth]{>}}},postaction={decorate}] (SB.south) to (NT.north);
	\draw [decoration={markings,mark=at position 1 with
    {\arrow[scale=3,>=stealth]{>}}},postaction={decorate}] (GWAE.south) to (-.5,-1.6);
    \draw [decoration={markings,mark=at position 1 with
    {\arrow[scale=3,>=stealth]{>}}},postaction={decorate}] (CS.south) to (FG.north)+(0,-1.3);
    \draw [decoration={markings,mark=at position 1 with
    {\arrow[scale=3,>=stealth]{>}}},postaction={decorate}] (NT.south) to (0.5,-1.6);
\end{tikzpicture}
\caption{Dependency structure between parameters, when final grades of the
students are observed.}
\label{fig:dep_grade}
\end{figure}

Here we assume the average size of a class and the number of the teachers at
the school are normally distributed with some mean, depending on the budget
of the school, and standard deviation equal to $1$.  We further assume that
the budget of the school is uniformly distributed between $1$ and $10$
millions US dollars.

We can define these random variables and their dependencies in \pkg{ABCpy}
in the following way:
%
\begin{CodeChunk}
\begin{CodeInput}
>>> from abcpy.continuousmodels import Uniform, Normal
>>> school_budget = Uniform([[1], [10]], name = "school_budget")
>>> class_size = Normal([[800*school_budget], [1]], name = "class_size")
>>> no_teacher = Normal([[20*school_budget], [1]], name = "no_teacher")
>>> historical_mean_grade = Normal([[4.5], [0.25]],
...   name = "historical_mean_grade")
\end{CodeInput}
\end{CodeChunk}
We model the impact of class size and the number of teachers on the final
grade each student receives in the following way:
%
\begin{CodeChunk}
\begin{CodeInput}
>>> final_grade = historical_mean_grade - .001 * class_size +
...   .02 * no_teacher
\end{CodeInput}
\end{CodeChunk}
Notice here we created a new random variable \code{final_grade}, by
subtracting the random variables \code{class\_size} and adding
\code{no_teacher}, suitably scaled, from the random variable
\code{historical_mean_grade}.  The resulting graphical model is represented
in Figure~\ref{fig:dep_grade}.

In short, this illustrates that you can {natively} perform standard
operations ``\code{+}'', ``\code{-}'', ``\code{*}'', ``\code{/}'' and
``\code{**}'' (the power operator in \proglang{Python}) on any two random
variables to get a new random variable.  It is possible to perform these
operations between random variables on top of the general data types of
\proglang{Python} (integer, float, and so on) since they are internally
converted to \code{HyperParameters}.  {If additional custom operations are
needed, users can implement those by sub-classing the
\class{ModelResultingFromOperation} class.}

\subsection{Co-occurring data set} \label{sec:hier_mod}

\pkg{ABCpy} supports inference when co-occurring (multiple) datasets are
available.  To illustrate how this is implemented, we extend the example
from Section~\ref{sec:prob_dep_rv} {to the case where }we also have data on
student with scholarships, stored in the variable \code{scholarship_obs}.
\begin{figure}
\centering
\begin{tikzpicture}
	\node[element_elip,text width=3cm]  (SB) at (1, 2) {School Budget};
	\node[element_elip,text width=2cm]  (CS) at (0, 0) {Class Size};
	\node[element_elip,text width=2.25cm]  (GWAE) at (-4, 0) {Historical mean grade};
	\node[element_elip,text width=2cm]  (NT) at (4, 0) {Number of teachers};
	\node[element_elip,text width=2.25cm]  (SWAE) at (8, 0) {Historical mean scholarship};
	\node[element_elip,text width=2.5cm]  (FG) at (0, -2) {Final Grade};
	\node[element_elip,text width=3.5cm]  (FS) at (6, -2.5) {Final Scholarship};
	
	\draw [decoration={markings,mark=at position 1 with
    {\arrow[scale=3,>=stealth]{>}}},postaction={decorate}] (SB.south) to (CS.north);
    	\draw [decoration={markings,mark=at position 1 with
    {\arrow[scale=3,>=stealth]{>}}},postaction={decorate}] (SB.south) to (NT.north);
	\draw [decoration={markings,mark=at position 1 with
    {\arrow[scale=3,>=stealth]{>}}},postaction={decorate}] (GWAE.south) to (-.5,-1.6);
    \draw [decoration={markings,mark=at position 1 with
    {\arrow[scale=3,>=stealth]{>}}},postaction={decorate}] (CS.south) to (FG.north)+(0,-1.3);
    \draw [decoration={markings,mark=at position 1 with
    {\arrow[scale=3,>=stealth]{>}}},postaction={decorate}] (NT.south) to (0.5,-1.6);
    \draw [decoration={markings,mark=at position 1 with
    {\arrow[scale=3,>=stealth]{>}}},postaction={decorate}] (NT.south) to (FS.north);
    \draw [decoration={markings,mark=at position 1 with
    {\arrow[scale=3,>=stealth]{>}}},postaction={decorate}] (SWAE.south) to (6.4,-2.1);
\end{tikzpicture}
\caption{Dependency structure between parameters, when final grades of the
students and their scholarship are observed.}
\label{fig:dep_grade_schol}
\end{figure}

We assume that the final mark of a student awarded a scholarship is similar
to the historical mean (restricted now to scholarship students), but there
is a correction dependent on the number of teachers in the school; we
therefore model it in the following way:
%
\begin{CodeChunk}
\begin{CodeInput}
>>> historical_mean_scholarship = Normal([[2], [0.5]],
...   name = "historical_mean_scholarship")
>>> final_scholarship = historical_mean_scholarship + .03 * no_teacher
\end{CodeInput}
\end{CodeChunk}
With this extension, we now have
two ``root'' \code{ProbabilisticModels} (random variables), namely
\code{final_grade} and \code{final_scholarship} (see
Figure~\ref{fig:dep_grade_schol}), whose output can be directly compared to
the observed datasets \code{grade_obs} and \code{scholarship_obs}.

Now, we need to choose summary statistics, distance, inference scheme,
backend and kernel.  However, since we are now considering two observed
datasets, we define statistics and distances on them separately.  In this
example, we use the \code{Identity} statistics (with different polynomial
expansion parameters \code{degree} and \code{cross}) and \code{Euclidean}
for both datasets, but in general these can be different:
%
\begin{CodeChunk}
\begin{CodeInput}
>>> # Define a summary statistics for final grade and final scholarship
>>> from abcpy.statistics import Identity
>>> statistics_final_grade = Identity(degree = 2, cross = False)
>>> statistics_final_scholarship = Identity(degree = 3, cross = False)

>>> # Define a distance measure for final grade and final scholarship
>>> from abcpy.distances import Euclidean
>>> distance_final_grade = Euclidean(statistics_final_grade)
>>> distance_final_scholarship = Euclidean(statistics_final_scholarship)

>>> # Define a backend
>>> from abcpy.backends import BackendDummy as Backend
>>> backend = Backend()

>>> # Define a perturbation kernel
>>> from abcpy.perturbationkernel import DefaultKernel
>>> kernel = DefaultKernel([school_budget, class_size,
...   historical_mean_grade, no_teacher, historical_mean_scholarship])

>>> # Define sampling parameters
>>> T, n_sample, n_samples_per_param = 3, 250, 10
>>> eps_arr = np.array([.75])
>>> eps_percentile = 10

>>> # Define sampler
>>> from abcpy.inferences import PMCABC
>>> sampler = PMCABC([final_grade, final_scholarship],
...   [distance_final_grade, distance_final_scholarship], backend, kernel)

>>> # Sample
>>> journal = sampler.sample([grades_obs, scholarship_obs], T, eps_arr,
...   n_sample, n_samples_per_param, eps_percentile)
\end{CodeInput}
\end{CodeChunk}
Notice that the lists passed to the sampler and the sampling method now
contain two entries, each corresponding to the different observed data sets
and models respectively.  Presently \pkg{ABCpy} combines different distances
on different datasets by taking an equally weighted convex linear
combination of the distances, however customized combination strategies can
be implemented by the user.
\newpage
\subsection{Joint perturbation kernels} \label{sec:comp_pert_ker}

As pointed out earlier, it is possible to define joint perturbation kernels,
perturbing different subsets of random variables using different kernel
functions.  Considering the example from Section~\ref{sec:hier_mod}, now we
want to perturb the schools budget, scholarship and grade variables using a
multivariate normal kernel, and we want to perturb the remaining parameters
with a multivariate student's-$T$ kernel.  This can be implemented as follows:
%
\begin{CodeChunk}
\begin{CodeInput}
>>> from abcpy.perturbationkernel import MultivariateNormalKernel,
...   MultivariateStudentTKernel
>>> kernel_1 = MultivariateNormalKernel([school_budget,
...   historical_mean_grade, historical_mean_scholarship])
>>> kernel_2 = MultivariateStudentTKernel([class_size, no_teacher], df = 3)

>>> # Join the defined kernels
>>> from abcpy.perturbationkernel import JointPerturbationKernel
>>> kernel = JointPerturbationKernel([kernel_1, kernel_2])
\end{CodeInput}
\end{CodeChunk}
In the last line, we use the class 
\class{abcpy.perturbationkernel.JointPerturbationKernel} to join the two
different kernels in a single one, by instantiating an object which takes as
parameters the kernels to join; this is needed as the sampler object needs
to be provided with one single kernel.  {Note that, in this way, it is
possible to combine kernels with dependency structures among disjoint
subsets of the parameters.}

As a side remark, note also that we cannot use the access operator to
perturb one component of a multidimensional random variable differently from
another component of the same variable.

\subsection{Nested parallelization}\label{sec:nested_par}

As mentioned above, \pkg{ABCpy} provides the user with seamless
parallelization of ABC algorithms using \pkg{MPI} or \pkg{Spark}.  Modern
cluster nodes have usually multiple cores, and by default {the \pkg{MPI}
backend} runs one simulation of the model per core.  Yet, in case the model
supports basic multi-threading at the level of a single machine, the backend
can be accordingly configured to achieve this.
	
There may be however cases in which simulation from the model is extremely
computationally demanding, so that each simulation has to be distributed
across different nodes at the same time of the parallel execution of
different simulations corresponding to different parameter values coming
from the use of ABC algorithm.  This is possible within \pkg{ABCpy} by using
the \pkg{MPI} backend.  Specifically, the model itself has to be implemented
with \pkg{MPI}, i.e.,~it has to be independently capable of running over
different nodes.  In this case, the \pkg{MPI} backend in \pkg{ABCpy}
controls the number of ranks that are assigned to each run of the model. 
For instance, consider defining the following backend for running
simulations on a cluster where each node has one single processor:
%
\begin{CodeChunk}
\begin{CodeInput}
>>> from abcpy.backends import BackendMPI as Backend
>>> backend = Backend(process_per_model = 2)
\end{CodeInput}
\end{CodeChunk}
As we require two ranks per model simulation, the \pkg{MPI} backend will
automatically split each model run on two different nodes.  We remark
instead that, if the number of cores in each node is larger than the
requested \code{process_per_model}, then the \pkg{MPI} backend will run each
simulation on two cores belonging to the same node.

\begin{figure}[t!]
	\centering
		\includegraphics[width = \textwidth]{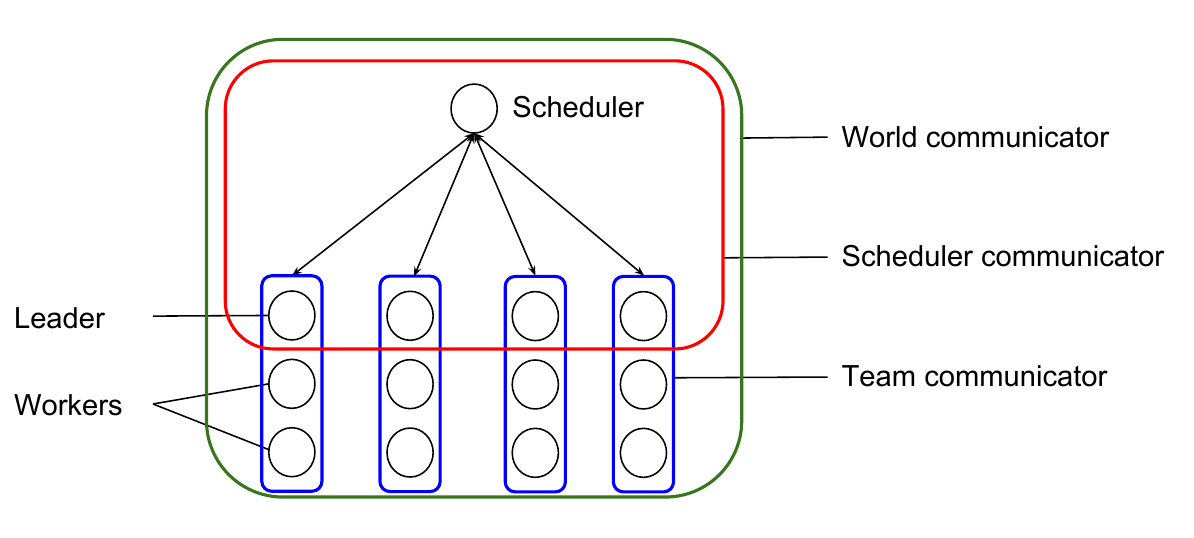}
	\caption{\textbf{Nested parallelization:} Description of the communication
	architecture of the nested \pkg{MPI} parallelization for \pkg{ABCpy}. Each circle
	represents a different rank.}
	\label{fig:nested_paral}
\end{figure}

Technically, \pkg{MPI} uses an object called \emph{communicator} in order to
control communication between different ranks.  Therefore, in order to
achieve nested parallelization \pkg{ABCpy} creates two kind of
communicators: each model simulation uses a \emph{team communicator} to
parallelize the computation on the ranks allocated by the \code{Backend}
object; moreover, the \emph{scheduler communicator} is used by the overall
master (the scheduler) to control the whole execution.  The architecture is
visualized in Figure~\ref{fig:nested_paral}.  Note that one process of each
team communicator is part of the scheduler one as well, in order for
communication to be successful.

More details on the nested parallelization scheme and an example of
successful application of ABC inference in such a scenario can be found in
\citet{pacchiardi2020distance}.

\subsection{Convergence diagnostic tools}\label{sec:convergence_checks}

Most algorithms implemented in \pkg{ABCpy} are SMC-type ones (particle
filtering); namely, these are sequential algorithms in which a set of
weighted particles represent an approximation of the target distribution and
are evolved across iterations.  As noted in \cite{del2007sequential},
contrarily to MCMC-type algorithms, SMC methods do not rely on ergodic
properties of the transition kernels.  For this reason, there is no need in
SMC to perform convergence checks of the kind used in MCMC for assessing
whether the chain convergent to the correct distribution.

However, the weights of the particles in these algorithms
(e.g.,~$\ith\omega_t$ for PMCABC in Algorithm~\ref{alg:PMCABC}) can
degenerate to a state in which all of the weight is attributed to one single
particle.  For this reason, it was suggested in \cite{del2007sequential} to
monitor the effective sample size (ESS) as the algorithm proceeds.  For a
set of $ n $ particles with normalized weights $\lbrace \ith\omega_t:
i=1,\ldots,n \rbrace$ at $t$-th iteration, the ESS at $t$-th iteration can
be estimated as:
\begin{equation*}\label{}
\text{ESS}_t = \left(\sum_{i=1}^n (\ith\omega_t)^2\right)^{-1};
\end{equation*}
the above reaches a maximum value of $ n $ when all weights are equal, and
can be as low as $ 1 $, when all the weight is borne by one single particle. 
In \pkg{ABCpy}, when ABC inference is performed with a sequential algorithm,
the ESS is computed at each iteration and stored.  Subsequently, it is
possible to produce a plot displaying its evolution with the following line:
\begin{CodeChunk}
\begin{CodeInput}
>>> journal.plot_ESS()
\end{CodeInput}
\end{CodeChunk}
Another possibility to assess the convergence of sequential algorithms is
computing some measure of distance between the set of samples at subsequent
iterations; in fact, the latter can be thought of as defining an empirical
distributions approximating the target one.  If the approximation is
converging, the change as the algorithm proceeds would become smaller.  In
\pkg{ABCpy}, we have implemented a tool to compute the 2-Wasserstein
distance \citep{peyre2019computational} between the distributions obtained
at subsequent iterations; such metric is chosen as it is a sensible measure
of distance between empirical distributions.  If the sequential algorithm
converges to some approximate distribution, we expect the Wasserstein
distance between subsequent iterations to decrease and become smaller.  In
\pkg{ABCpy}, the evolution of the Wasserstein distance can be plot with the
following command:
\begin{CodeChunk}
\begin{CodeInput}
>>> journal.Wass_convergence_plot()
\end{CodeInput}
\end{CodeChunk}

\section{Discussion} \label{sec:conclusion} 

There has been significant interest and efforts to develop new algorithms
for ABC.  A timely need in this area is to create an ecology where all these
different algorithms can be integrated in a modular and user-friendly
manner.  It is also known that ABC algorithms can be very expensive and
without HPC integration they cannot be applied to computationally intensive
simulator-based models.  Although the SMCABC algorithm had been parallelized
before \citep{Liepe_2010}, more efficient algorithms have since then been
suggested (for instance, SABC in \citet{Albert_2015}).  It is therefore very
important to provide a simple way to parallelize ABC algorithms within an
unified ecology and compare their parallel performance.

Our main contribution is a framework that (i) brings existing ABC algorithms
under one umbrella, (ii) enables easy implementation of new ABC algorithms,
and (iii) enables domain scientists to easily apply ABC to their specific
problem on a broad scale using parallelization.  For point (i), it is
important to note that, although there is a strong current interest in ABC,
there are only a few software libraries available and, up to our knowledge,
none, concurrently, as complete, user-friendly, and extensible as
\pkg{ABCpy}.  To add to point (ii), we stress that having a unified,
extensible library is one of the foundations of a principled and
reproducible comparison of algorithms.  In this paper, we provide a
comparison of ABC algorithms from a parallel performance perspective.  Hence
we have reported on imbalances while parallelizing ABC type algorithms over
a large number of cores.  We identified inherent properties of ABC
algorithms that make efficient parallelization difficult, classified ABC
algorithms based on the imbalances, and tried to find the most suitable
algorithms capable of utilizing a large parallel architecture through
empirical comparisons.

\section*{Acknowledgments}

The work was supported by Swiss National Science Foundation Grant No. 
105218\_163196 (Statistical Inference on Large-Scale Mechanistic Network
Models).  We also thank Swiss National Super Computing Center for providing
computing resources.  LP is supported by the EPSRC and MRC through the
OxWaSP CDT program (EP/L016710/1).
 

\begin{thebibliography}{60}
\newcommand{\enquote}[1]{``#1''}
\providecommand{\natexlab}[1]{#1}
\providecommand{\url}[1]{\texttt{#1}}
\providecommand{\urlprefix}{URL }
\expandafter\ifx\csname urlstyle\endcsname\relax
  \providecommand{\doi}[1]{doi:\discretionary{}{}{}#1}\else
  \providecommand{\doi}{doi:\discretionary{}{}{}\begingroup
  \urlstyle{rm}\Url}\fi
\providecommand{\eprint}[2][]{\url{#2}}

\bibitem[{Albert \emph{et~al.}(2015)Albert, Künsch, and
  Scheidegger}]{Albert_2015}
Albert C, Künsch HR, Scheidegger A (2015).
\newblock \enquote{A Simulated Annealing Approach to Approximate {B}ayesian
  Computations.}
\newblock \emph{Statistics and Computing}, \textbf{25}, 1217--1232.
\newblock \doi{10.1007/s11222-014-9507-8}.

\bibitem[{Amdahl(1967)}]{amdahl1967validity}
Amdahl GM (1967).
\newblock \enquote{Validity of the Single Processor Approach to Achieving Large
  Scale Computing Capabilities.}
\newblock In \emph{Proceedings of the April 18--20, 1967, Spring Joint Computer
  Conference}, pp. 483--485. ACM.

\bibitem[{An \emph{et~al.}(2020)An, Nott, and Drovandi}]{an2020robust}
An Z, Nott DJ, Drovandi C (2020).
\newblock \enquote{Robust {B}ayesian Synthetic Likelihood via a Semi-Parametric
  Approach.}
\newblock \emph{Statistics and Computing}, \textbf{30}(3), 543--557.
\newblock \doi{10.1007/s11222-019-09904-x}.

\bibitem[{Andrieu and Roberts(2009)}]{Andrieu_2009}
Andrieu C, Roberts GO (2009).
\newblock \enquote{The Pseudo-Marginal Approach for Efficient {M}onte {C}arlo
  Computations.}
\newblock \emph{The Annals of Statistics}, \textbf{37}(2), 697--725.
\newblock \doi{10.1214/07-aos574}.

\bibitem[{Beaumont(2010)}]{Beaumont2010}
Beaumont MA (2010).
\newblock \enquote{Approximate {B}ayesian Computation in Evolution and
  Ecology.}
\newblock \emph{Annual Review of Ecology, Evolution, and Systematics},
  \textbf{41}(1), 379--406.
\newblock \doi{10.1146/annurev-ecolsys-102209-144621}.

\bibitem[{Beaumont \emph{et~al.}(2002)Beaumont, Zhang, and
  Balding}]{Beaumont2002}
Beaumont MA, Zhang W, Balding DJ (2002).
\newblock \enquote{Approximate {B}ayesian Computation in Population Genetics.}
\newblock \emph{Genetics}, \textbf{162}(4), 2025--2035.
\newblock \doi{10.1093/genetics/162.4.2025}.

\bibitem[{Beazley \emph{et~al.}(1996)}]{beazley1996swig}
Beazley DM, \emph{et~al.} (1996).
\newblock \enquote{\pkg{SWIG}: An Easy to Use Tool for Integrating Scripting
  Languages with \proglang{C} and \proglang{C++.}}
\newblock In \emph{Tcl/Tk Workshop}, volume~43, p.~74.

\bibitem[{Bernton \emph{et~al.}(2019)Bernton, Jacob, Gerber, and
  Robert}]{bernton2019}
Bernton E, Jacob PE, Gerber M, Robert CP (2019).
\newblock \enquote{Approximate Bayesian Computation with the Wasserstein
  Distance.}
\newblock \emph{Journal of the Royal Statistical Society B}, \textbf{81}(2),
  235--269.
\newblock \doi{10.1111/rssb.12312}.

\bibitem[{Blum and François(2010)}]{blum2010non}
Blum MG, François O (2010).
\newblock \enquote{Non-Linear Regression Models for Approximate Bayesian
  Computation.}
\newblock \emph{Statistics and Computing}, \textbf{20}(1), 63--73.
\newblock \doi{10.1007/s11222-009-9116-0}.

\bibitem[{Cappé \emph{et~al.}(2004)Cappé, Guillin, Marin, and
  Robert}]{Cappe_2004}
Cappé O, Guillin A, Marin JM, Robert CP (2004).
\newblock \enquote{Population {M}onte {C}arlo.}
\newblock \emph{Journal of Computational and Graphical Statistics},
  \textbf{13}(4), 907--929.
\newblock \doi{10.1198/106186004x12803}.

\bibitem[{Carlson(2013)}]{redis}
Carlson JL (2013).
\newblock \emph{\pkg{Redis} in Action}.
\newblock Manning Publications Co., USA.

\bibitem[{Chiachio \emph{et~al.}(2014)Chiachio, Beck, Chiachio, and
  Rus}]{Chiachio_2014}
Chiachio M, Beck JL, Chiachio J, Rus G (2014).
\newblock \enquote{Approximate {B}ayesian Computation by Subset Simulation.}
\newblock \emph{SIAM Journal on Scientific Computing}, \textbf{36}(3),
  A1339--A1358.
\newblock \doi{10.1137/130932831}.

\bibitem[{Csilléry \emph{et~al.}(2012)Csilléry, François, and
  Blum}]{csillery2012abc}
Csilléry K, François O, Blum MG (2012).
\newblock \enquote{\pkg{abc}: An \proglang{R} Package for Approximate Bayesian
  Computation (ABC).}
\newblock \emph{Methods in Ecology and Evolution}, \textbf{3}(3), 475--479.
\newblock \doi{10.1111/j.2041-210x.2011.00179.x}.

\bibitem[{{\pkg{Dask} Development Team}(2016)}]{dask}
{\pkg{Dask} Development Team} (2016).
\newblock \emph{\pkg{Dask}: Library for Dynamic Task Scheduling}.
\newblock \urlprefix\url{https://dask.org}.

\bibitem[{Dean and Ghemawat(2008)}]{Dean2008Mapreduce}
Dean J, Ghemawat S (2008).
\newblock \enquote{MapReduce: Simplified Data Processing on Large Clusters.}
\newblock \emph{Communications of the ACM}, \textbf{51}(1), 107--113.
\newblock \doi{10.1145/1327452.1327492}.

\bibitem[{{Del Moral} \emph{et~al.}(2007){Del Moral}, Doucet, and
  Jasra}]{del2007sequential}
{Del Moral} P, Doucet A, Jasra A (2007).
\newblock \enquote{Sequential Monte Carlo for Bayesian Computation.}
\newblock \emph{Bayesian Statistics}, \textbf{8}, 1--34.
\newblock \doi{10.1007/s11222-011-9271-y}.

\bibitem[{{Del Moral} \emph{et~al.}(2012){Del Moral}, Doucet, and
  Jasra}]{del2012adaptive}
{Del Moral} P, Doucet A, Jasra A (2012).
\newblock \enquote{An Adaptive Sequential {M}onte {C}arlo Method for
  Approximate {B}ayesian Computation.}
\newblock \emph{Statistics and Computing}, \textbf{22}(5), 1009--1020.
\newblock \doi{10.1007/s11222-011-9271-y}.

\bibitem[{Drovandi and Pettitt(2011)}]{drovandi2011estimation}
Drovandi CC, Pettitt AN (2011).
\newblock \enquote{Estimation of Parameters for Macroparasite Population
  Evolution Using Approximate {B}ayesian Computation.}
\newblock \emph{Biometrics}, \textbf{67}(1), 225--233.
\newblock \doi{10.1111/j.1541-0420.2010.01410.x}.

\bibitem[{Fearnhead and Prangle(2012)}]{fearnhead2012constructing}
Fearnhead P, Prangle D (2012).
\newblock \enquote{Constructing Summary Statistics for Approximate Bayesian
  Computation: Semi-Automatic Approximate Bayesian Computation.}
\newblock \emph{Journal of the Royal Statistical Society B}, \textbf{74}(3),
  419--474.
\newblock \doi{10.1111/j.1467-9868.2011.01010.x}.

\bibitem[{Graham(1966)}]{Graham1966}
Graham RL (1966).
\newblock \enquote{Bounds for Certain Multiprocessing Anomalies.}
\newblock \emph{Bell Labs Technical Journal}, \textbf{45}(9), 1563--1581.
\newblock \doi{10.1002/j.1538-7305.1966.tb01709.x}.

\bibitem[{Gutmann and Corander(2016)}]{gutmann2016bayesian}
Gutmann MU, Corander J (2016).
\newblock \enquote{Bayesian Optimization for Likelihood-Free Inference of
  Simulator-Based Statistical Models.}
\newblock \emph{The Journal of Machine Learning Research}, \textbf{17}(1),
  4256--4302.

\bibitem[{Gutmann \emph{et~al.}(2018)Gutmann, Dutta, Kaski, and
  Corander}]{Gutmann_2014}
Gutmann MU, Dutta R, Kaski S, Corander J (2018).
\newblock \enquote{Likelihood-Free Inference via Classification.}
\newblock \emph{Statistics and Computing}, \textbf{28}, 411--425.
\newblock \doi{10.1007/s11222-017-9738-6}.

\bibitem[{Hakkarainen \emph{et~al.}(2012)Hakkarainen, Ilin, Solonen, Laine,
  Haario, Tamminen, Oja, and Järvinen}]{Hakkarainen_2012}
Hakkarainen J, Ilin A, Solonen A, Laine M, Haario H, Tamminen J, Oja E,
  Järvinen H (2012).
\newblock \enquote{{O}n Closure Parameter Estimation in Chaotic Systems.}
\newblock \emph{Nonlinear Processes in Geophysics}, \textbf{19}(1), 127--143.
\newblock \doi{10.5194/npg-19-127-2012}.

\bibitem[{Harris \emph{et~al.}(2020)Harris, Millman, {van der Walt}, Gommers,
  Virtanen, Cournapeau, Wieser, Taylor, Berg, Smith, Kern, Picus, Hoyer, {van
  Kerkwijk}, Brett, Haldane, {del Río}, Wiebe, Peterson, Gérard-Marchant,
  Sheppard, Reddy, Weckesser, Abbasi, Gohlke, and Oliphant}]{harris2020array}
Harris CR, Millman KJ, {van der Walt} SJ, Gommers R, Virtanen P, Cournapeau D,
  Wieser E, Taylor J, Berg S, Smith NJ, Kern R, Picus M, Hoyer S, {van
  Kerkwijk} MH, Brett M, Haldane A, {del Río} JF, Wiebe M, Peterson P,
  Gérard-Marchant P, Sheppard K, Reddy T, Weckesser W, Abbasi H, Gohlke C,
  Oliphant TE (2020).
\newblock \enquote{Array Programming with \pkg{NumPy}.}
\newblock \emph{Nature}, \textbf{585}(7825), 357--362.
\newblock \doi{10.1038/s41586-020-2649-2}.

\bibitem[{Hastings(1970)}]{Hastings}
Hastings WK (1970).
\newblock \enquote{Monte Carlo Sampling Methods Using Markov Chains and Their
  Applications.}
\newblock \emph{Biometrika}, \textbf{57}(1), 97--109.
\newblock \doi{10.1093/biomet/57.1.97}.

\bibitem[{{\pkg{IPython} Development Team}(2021)}]{ipyparallel}
{\pkg{IPython} Development Team} (2021).
\newblock \emph{\pkg{ipyparallel} Docs}.
\newblock \urlprefix\url{https://github.com/ipython/ipyparallel}.

\bibitem[{Jabot \emph{et~al.}(2015)Jabot, Faure, Dumoulin, and
  Albert}]{jaboteasyabc}
Jabot F, Faure T, Dumoulin N, Albert C (2015).
\newblock \emph{\pkg{EasyABC}: Efficient Approximate Bayesian Computation
  Sampling Schemes}.
\newblock \proglang{R}~package version~1.5,
  \urlprefix\url{https://CRAN.R-project.org/package=EasyABC}.

\bibitem[{Jennings and Madigan(2017)}]{Jennings_2016}
Jennings E, Madigan M (2017).
\newblock \enquote{\pkg{astro{ABC}}: An Approximate {B}ayesian Computation
  Sequential {M}onte {C}arlo Sampler for Cosmological Parameter Estimation.}
\newblock \emph{Astronomy and Computing}, \textbf{19}, 16--22.
\newblock \doi{10.1016/j.ascom.2017.01.001}.

\bibitem[{Jiang \emph{et~al.}(2017)Jiang, Wu, Zheng, and
  Wong}]{jiang2017learning}
Jiang B, Wu T, Zheng C, Wong WH (2017).
\newblock \enquote{Learning Summary Statistic for Approximate Bayesian
  Computation via Deep Neural Network.}
\newblock \emph{Statistica Sinica}, pp. 1595--1618.
\newblock \doi{10.5705/ss.202015.0340}.

\bibitem[{Klinger \emph{et~al.}(2018)Klinger, Rickert, and
  Hasenauer}]{klinger2018pyabc}
Klinger E, Rickert D, Hasenauer J (2018).
\newblock \enquote{\pkg{pyABC}: Distributed, Likelihood-Free Inference.}
\newblock \emph{Bioinformatics}, \textbf{34}(20), 3591--3593.
\newblock \doi{10.1093/bioinformatics/bty361}.

\bibitem[{Kulakova \emph{et~al.}(2016)Kulakova, Angelikopoulos, Hadjidoukas,
  Papadimitriou, and Koumoutsakos}]{Kulakova_2016}
Kulakova L, Angelikopoulos P, Hadjidoukas PE, Papadimitriou C, Koumoutsakos P
  (2016).
\newblock \enquote{Approximate {B}ayesian Computation for Granular and
  Molecular Dynamics Simulations.}
\newblock In \emph{Proceedings of the Platform for Advanced Scientific
  Computing Conference}, PASC '16, pp. 4:1--4:12. ACM.
\newblock \doi{10.1145/2929908.2929918}.

\bibitem[{Lenormand \emph{et~al.}(2013)Lenormand, Jabot, and
  Deffuant}]{lenormand2013adaptive}
Lenormand M, Jabot F, Deffuant G (2013).
\newblock \enquote{Adaptive Approximate {B}ayesian Computation for Complex
  Models.}
\newblock \emph{Computational Statistics}, \textbf{28}(6), 2777--2796.
\newblock \doi{10.1007/s00180-013-0428-3}.

\bibitem[{Liepe \emph{et~al.}(2010)Liepe, Barnes, Cule, Erguler, Kirk, Toni,
  and Stumpf}]{Liepe_2010}
Liepe J, Barnes C, Cule E, Erguler K, Kirk P, Toni T, Stumpf MPH (2010).
\newblock \enquote{{\pkg{{ABC-S}ys{B}io} -- Approximate {B}ayesian Computation
  in \proglang{Python} with {GPU} Support}.}
\newblock \emph{Bioinformatics}, \textbf{26}(14), 1797--1799.
\newblock \doi{10.1093/bioinformatics/btq278}.

\bibitem[{Lintusaari \emph{et~al.}(2016)Lintusaari, Gutmann, Dutta, Kaski, and
  Corander}]{Lintusaari_2016}
Lintusaari J, Gutmann MU, Dutta R, Kaski S, Corander J (2016).
\newblock \enquote{Fundamentals and Recent Developments in Approximate
  {B}ayesian Computation.}
\newblock \emph{Systematic Biology}, \textbf{66}(1), e66--e82.
\newblock \doi{10.1093/sysbio/syw077}.

\bibitem[{Lintusaari \emph{et~al.}(2018)Lintusaari, Vuollekoski,
  Kangasrääsiö, Skytén, Järvenpää, Marttinen, Gutmann, Vehtari,
  Corander, and Kaski}]{lintusaari2018elfi}
Lintusaari J, Vuollekoski H, Kangasrääsiö A, Skytén K, Järvenpää M,
  Marttinen P, Gutmann MU, Vehtari A, Corander J, Kaski S (2018).
\newblock \enquote{\pkg{ELFI}: Engine for Likelihood-Free Inference.}
\newblock \emph{The Journal of Machine Learning Research}, \textbf{19}(1),
  643--649.

\bibitem[{Lorenz(1995)}]{Lorenz_1995}
Lorenz EN (1995).
\newblock \enquote{Predictability: A Problem Partly Solved.}
\newblock In \emph{Proceedings of the Seminar on Predictability, 4-8 September
  1995}, volume~1, pp. 1--18. European Center on Medium Range Weather
  Forecasting, European Center on Medium Range Weather Forecasting, Shinfield
  Park, Reading.

\bibitem[{Marin \emph{et~al.}(2012)Marin, Pudlo, Robert, and
  Ryder}]{Marin_2012}
Marin JM, Pudlo P, Robert C, Ryder R (2012).
\newblock \enquote{Approximate {B}ayesian Computational Methods.}
\newblock \emph{Statistics and Computing}, \textbf{22}(6), 1167--1180.
\newblock \doi{10.1007/s11222-011-9288-2}.

\bibitem[{Martinez \emph{et~al.}(2016)Martinez, Muschik, Schindler, Nigg,
  Erhard, Heyl, Hauke, Dalmonte, Monz, Zoller, and Blatt}]{Martinez_2016}
Martinez EA, Muschik CA, Schindler P, Nigg D, Erhard A, Heyl M, Hauke P,
  Dalmonte M, Monz T, Zoller P, Blatt R (2016).
\newblock \enquote{Real-Time Dynamics of Lattice Gauge Theories with a
  Few-Qubit Quantum Computer.}
\newblock \emph{Nature}, \textbf{534}(7608), 516--519.
\newblock \doi{10.1038/nature18318}.

\bibitem[{{Message Passing Interface Forum}(2012)}]{message2012mpi}
{Message Passing Interface Forum} (2012).
\newblock \emph{\pkg{MPI}: A Message Passing Interface Standard}.

\bibitem[{Pacchiardi and Dutta(2020)}]{pacchiardi2020score}
Pacchiardi L, Dutta R (2020).
\newblock \enquote{Score Matched Conditional Exponential Families for
  Likelihood-Free Inference.}
\newblock {arXiv}:2012.10903 [stat.ME],
  \urlprefix\url{https://arxiv.org/abs/2012.10903}.

\bibitem[{Pacchiardi \emph{et~al.}(2020)Pacchiardi, Künzli, Schöngens,
  Chopard, and Dutta}]{pacchiardi2020distance}
Pacchiardi L, Künzli P, Schöngens M, Chopard B, Dutta R (2020).
\newblock \enquote{Distance-Learning for Approximate Bayesian Computation to
  Model a Volcanic Eruption.}
\newblock \emph{Sankhya B}, pp. 1--30.
\newblock \doi{10.1007/s13571-019-00208-8}.

\bibitem[{Paszke \emph{et~al.}(2017)Paszke, Gross, Chintala, Chanan, Yang,
  DeVito, Lin, Desmaison, Antiga, and Lerer}]{paszke2017automatic}
Paszke A, Gross S, Chintala S, Chanan G, Yang E, DeVito Z, Lin Z, Desmaison A,
  Antiga L, Lerer A (2017).
\newblock \enquote{Automatic Differentiation in \pkg{PyTorch}.}
\newblock In \emph{NIPS Autodiff Workshop}.

\bibitem[{Peyré and Cuturi(2019)}]{peyre2019computational}
Peyré G, Cuturi M (2019).
\newblock \enquote{Computational Optimal Transport: With Applications to Data
  Science.}
\newblock \emph{Foundations and Trends{\textregistered} in Machine Learning},
  \textbf{11}(5-6), 355--607.
\newblock \doi{10.1561/2200000073}.

\bibitem[{Pritchard \emph{et~al.}(1999)Pritchard, Seielstad, Perez-Lezaun, and
  Feldman}]{Pritchard1999}
Pritchard JK, Seielstad MT, Perez-Lezaun A, Feldman MW (1999).
\newblock \enquote{Population Growth of Human {Y} Chromosomes: A Study of {Y}
  Chromosome Microsatellites.}
\newblock \emph{Molecular Biology and Evolution}, \textbf{16}(12), 1791--1798.
\newblock \doi{10.1093/oxfordjournals.molbev.a026091}.

\bibitem[{Pudlo \emph{et~al.}(2015)Pudlo, Marin, Estoup, Cornuet, Gautier, and
  Robert}]{Pudlo_2015}
Pudlo P, Marin JM, Estoup A, Cornuet JM, Gautier M, Robert CP (2015).
\newblock \enquote{Reliable {ABC} Model Choice via Random Forests.}
\newblock \emph{Bioinformatics}, \textbf{32}(6), 859--866.
\newblock \doi{10.1093/bioinformatics/btv684}.

\bibitem[{Pérez and Granger(2007)}]{ipython}
Pérez F, Granger BE (2007).
\newblock \enquote{\pkg{IPython}: A System for Interactive Scientific
  Computing.}
\newblock \emph{Computing in Science and Engineering}, \textbf{9}(3), 21--29.
\newblock \doi{10.1109/mcse.2007.53}.

\bibitem[{{\proglang{R}~Core Team}(2021)}]{R}
{\proglang{R}~Core Team} (2021).
\newblock \emph{\proglang{R}: A Language and Environment for Statistical
  Computing}.
\newblock \proglang{R}~Foundation for Statistical Computing, Vienna, Austria.
\newblock \urlprefix\url{https://www.R-project.org/}.

\bibitem[{Robert and Casella(2005)}]{Robert2005}
Robert CP, Casella G (2005).
\newblock \emph{Monte Carlo Statistical Methods}.
\newblock Springer-Verlag.

\bibitem[{Schaye \emph{et~al.}(2015)Schaye, Crain, Bower, Furlong, Schaller,
  Theuns, {Dalla Vecchia}, Frenk, McCarthy, Helly, Jenkins, Rosas-Guevara,
  White, Baes, Booth, Camps, Navarro, Qu, Rahmati, Sawala, Thomas, and
  Trayford}]{Schaye_2015}
Schaye J, Crain RA, Bower RG, Furlong M, Schaller M, Theuns T, {Dalla Vecchia}
  C, Frenk CS, McCarthy IG, Helly JC, Jenkins A, Rosas-Guevara YM, White SDM,
  Baes M, Booth CM, Camps P, Navarro JF, Qu Y, Rahmati A, Sawala T, Thomas PA,
  Trayford J (2015).
\newblock \enquote{The EAGLE Project: Simulating the Evolution and Assembly of
  Galaxies and Their Environments.}
\newblock \emph{Monthly Notices of the Royal Astronomical Society},
  \textbf{446}(1), 521--554.
\newblock \doi{10.1093/mnras/stu2058}.

\bibitem[{Stram \emph{et~al.}(2015)Stram, Marjoram, and Chen}]{Stram_2015}
Stram AH, Marjoram P, Chen GK (2015).
\newblock \enquote{{\pkg{al3c}: High-Performance Software for Parameter
  Inference Using {A}pproximate {B}ayesian {C}omputation}.}
\newblock \emph{Bioinformatics}, \textbf{31}(21), 3549--3551.
\newblock \doi{10.1093/bioinformatics/btv393}.

\bibitem[{Tavaré \emph{et~al.}(1997)Tavaré, Balding, Griffiths, and
  Donnelly}]{Tavare1997}
Tavaré S, Balding DJ, Griffiths RC, Donnelly P (1997).
\newblock \enquote{Inferring Coalescence Times From {DNA} Sequence Data.}
\newblock \emph{Genetics}, \textbf{145}(2), 505--518.
\newblock \doi{10.1093/genetics/145.2.505}.

\bibitem[{Thomas \emph{et~al.}(2021)Thomas, Dutta, Corander, Kaski, and
  Gutmann}]{thomas2016likelihood}
Thomas O, Dutta R, Corander J, Kaski S, Gutmann MU (2021).
\newblock \enquote{Likelihood-Free Inference by Ratio Estimation.}
\newblock \emph{Bayesian Analysis}, pp. 1--31.
\newblock \doi{10.1214/20-ba1238}.

\bibitem[{Toni \emph{et~al.}(2009)Toni, Welch, Strelkowa, Ipsen, and
  Stumpf}]{Toni_2009}
Toni T, Welch D, Strelkowa N, Ipsen A, Stumpf MPH (2009).
\newblock \enquote{Approximate {B}ayesian Computation Scheme for Parameter
  Inference and Model Selection in Dynamical Systems.}
\newblock \emph{Journal of the Royal Society Interface}, \textbf{31}(6),
  187--202.
\newblock \doi{10.1098/rsif.2008.0172}.

\bibitem[{Turchin \emph{et~al.}(2013)Turchin, Currie, Turner, and
  Gavrilets}]{Turchin_2013}
Turchin P, Currie TE, Turner EAL, Gavrilets S (2013).
\newblock \enquote{War, Space, and the Evolution of Old World Complex
  Societies.}
\newblock \emph{Proceedings of the National Academy of Sciences of the United
  States of America}, \textbf{110}(41), 16384--16389.
\newblock \doi{10.1073/pnas.1308825110}.

\bibitem[{{van Rossum} \emph{et~al.}(2011)}]{python}
{van Rossum} G, \emph{et~al.} (2011).
\newblock \emph{\proglang{Python} Programming Language}.
\newblock \urlprefix\url{http://www.python.org}.

\bibitem[{Virtanen \emph{et~al.}(2020)Virtanen, Gommers, Oliphant, Haberland,
  Reddy, Cournapeau, Burovski, Peterson, Weckesser, Bright, {van der Walt},
  Brett, Wilson, Millman, Mayorov, Nelson, Jones, Kern, Larson, Carey, İlhan
  Polat, Feng, Moore, VanderPlas, Laxalde, Perktold, Cimrman, Henriksen,
  Quintero, Harris, Archibald, Ribeiro, Pedregosa, {van Mulbregt}, and
  {\pkg{SciPy} 1.0 Contributors}}]{2020SciPy-NMeth}
Virtanen P, Gommers R, Oliphant TE, Haberland M, Reddy T, Cournapeau D,
  Burovski E, Peterson P, Weckesser W, Bright J, {van der Walt} SJ, Brett M,
  Wilson J, Millman KJ, Mayorov N, Nelson ARJ, Jones E, Kern R, Larson E, Carey
  CJ, İlhan Polat, Feng Y, Moore EW, VanderPlas J, Laxalde D, Perktold J,
  Cimrman R, Henriksen I, Quintero EA, Harris CR, Archibald AM, Ribeiro AH,
  Pedregosa F, {van Mulbregt} P, {\pkg{SciPy} 10 Contributors} (2020).
\newblock \enquote{{\pkg{SciPy}~1.0: Fundamental Algorithms for Scientific
  Computing in \proglang{Python}}.}
\newblock \emph{Nature Methods}, \textbf{17}, 261--272.
\newblock \doi{10.1038/s41592-019-0686-2}.

\bibitem[{Wilks(2005)}]{Wilks_2005}
Wilks DS (2005).
\newblock \enquote{Effects of Stochastic Parametrizations in the {L}orenz '96
  System.}
\newblock \emph{Quarterly Journal of the Royal Meteorological Society},
  \textbf{131}(606), 389--407.
\newblock \doi{10.1256/qj.04.03}.

\bibitem[{Wiqvist \emph{et~al.}(2019)Wiqvist, Mattei, Picchini, and
  Frellsen}]{wiqvist2019partially}
Wiqvist S, Mattei PA, Picchini U, Frellsen J (2019).
\newblock \enquote{Partially Exchangeable Networks and Architectures for
  Learning Summary Statistics in Approximate {B}ayesian Computation.}
\newblock In K~Chaudhuri, R~Salakhutdinov (eds.), \emph{Proceedings of the 36th
  International Conference on Machine Learning}, volume~97 of \emph{Proceedings
  of Machine Learning Research}, pp. 6798--6807. PMLR.
\newblock \urlprefix\url{http://proceedings.mlr.press/v97/wiqvist19a.html}.

\bibitem[{Wood(2010)}]{Wood_2010}
Wood SN (2010).
\newblock \enquote{Statistical Inference for Noisy Nonlinear Ecological Dynamic
  Systems.}
\newblock \emph{Nature}, \textbf{466}(7310), 1102--1104.
\newblock \doi{10.1038/nature09319}.

\bibitem[{Zaharia \emph{et~al.}(2016)Zaharia, Xin, Wendell, Das, Armbrust,
  Dave, Meng, Rosen, Venkataraman, Franklin, Ghodsi, Gonzalez, Shenker, and
  Stoica}]{ApacheSpark}
Zaharia M, Xin RS, Wendell P, Das T, Armbrust M, Dave A, Meng X, Rosen J,
  Venkataraman S, Franklin MJ, Ghodsi A, Gonzalez J, Shenker S, Stoica I
  (2016).
\newblock \enquote{\pkg{Apache Spark}: A Unified Engine for Big Data
  Processing.}
\newblock \emph{Communications of the ACM}, \textbf{59}(11), 56--65.
\newblock \doi{10.1145/2934664}.

\end{thebibliography}

\newpage
\begin{appendix}
	
\section{Additional details on parameters inference for the Lorenz95 model}
\label{app:lorenz}

We give here more details on the Lorenz95 model considered as a running
example throughout the main text.  We used a modification of the original
weather prediction model of \cite{Lorenz_1995} when fast climate variables
are unobserved \cite{Wilks_2005}.
\begin{itemize}
\item \textbf{Model}: We assume that weather stations measure a
high-dimensional time-series of slow climate variables $(y_k^{(t)},
k=1,\ldots,40)$, following a coupled stochastic differential equation (SDE),
called the forecast model \citep{Wilks_2005}:
\begin{align}
\label{eq:Lorenz}
 &\frac{dy^{(t)}_k}{dt}=-y^{(t)}_{k-1}(y^{(t)}_{k-2}-y^{(t)}_{k+1})-y_k^{(t)}+F
  -g(y_k^{(t)},\theta)+\eta_{k}^{(t)}, \\
  &g(y_k^{(t)},\theta) = \sum_{i=1}^2\theta_i\left(y_k^{(t)}\right)^{i-1},
\end{align}
for $k = 1,\ldots,40$ and where $F=10$.  Assuming that the initial values
$y_k^{(0)}, k=1,\ldots, 40$ are known, we consider the interval $[0,4]$ in
the time units of the model.  The function $g(y_k^{(t)},\theta)$ represents
a deterministic parametrization of the net effect of the unobserved fast
weather variables on the observable $y_k^{(t)}$, and $\eta_{k}^{(t)}$ is a
stochastic forcing term representing the uncertainty due to the forcing of
the fast variables.  The model is cyclic in the variables $y^{(t)}_k$, and
the coupled SDEs do not have an analytic solution.

We discretize the time-interval $[0,4]$ into $T$ equal steps of length
$\Delta t = 4/T$, and solve the SDEs by using a 4th order Runge-Kutta solver
at these time-points.  Following \cite{Wilks_2005} the stochastic forcing
term is updated for an interval of $\Delta t$ as
\begin{eqnarray*}
 \eta_{k}^{(t+\Delta t)} &= \phi \eta_{k}^{(t)} + (1-\phi^2)^{\frac{1}{2}}e^{(t)},
   t \in \{0, \Delta t,  \ldots, T\Delta t\}
\label{eq:stoch_forceterm_Lorenz}
\end{eqnarray*}
where the $e^{(t)}$ are independent normal random variables with standard
deviation $\sigma_e$ and $\eta^{(0)} = (1-\phi^2)^{\frac{1}{2}}e^{(0)}$. 
Here $T$ is chosen to be 1024.

\item \textbf{Parameters}: We fix $ \phi = 0.4,\ \sigma_e=1 $ and infer the
parameters $\theta = (\theta_1,\theta_2)$.

\item \textbf{Prior}: We assume uniform prior distributions with ranges
$[0.5, 3.5]$ and $[0,0.3]$ for the parameters $\theta_1$ and $\theta_2$,
respectively; this is motivated by the observations in
\cite{Hakkarainen_2012}.

\item \textbf{Observed dataset ($\dataObs$)}: A multivariate time series
computed by solving the SDEs numerically, as described above, with $\theta^0
= (\theta^o_1,\theta^o_2)=(2.0,0.1)$ over a period $ t \in [0,4] $ with $
T=1024 $.

\item \textbf{Statistics}: The six summary statistics suggested by
\cite{Hakkarainen_2012}: for each $ k $, we compute the mean, variance and
auto-co-variance with time lag one of $y^{(t)}_k$, co-variance of
$y^{(t)}_k$ with its neighbor $y^{(t)}_{k+1}$ and cross-co-variance of
$y^{(t)}_k$ with its two neighbors $y^{(t)}_{k-1}$ and $y^{(t)}_{k+1}$ for
time lag one.  These values are all averaged over $k=1, \ldots, 40$ since
the model is symmetric with respect to the index $k$.

\item \textbf{Distance}: Euclidean distance in both the experiments with
hand-chosen and learned statistics.

\item \textbf{Experimental setting}: All of the algorithms considered in the
main text (PMCABC, APMCABC, SABC and ABCsubsim)  are sequential population
algorithms.  We run all of them for 20 steps (expect for PMCABC that is run
for only 3 steps, for the reasons described in
Section~\ref{sec:imbalance_classify_abc}) and drew 10,000 samples at each
step.  Therefore, at the end we are provided with 10,000 samples from the
approximate posterior distribution of the parameters.  A multivariate Normal
distribution with 3 degrees of freedom was used as the perturbation kernel
and the Euclidean distance as the discrepancy measure.  For the PMCABC
algorithm, we chose an initial threshold value $\epsilon=500$ for the first
step of the algorithm.  For the subsequent steps, the $0.1$-quantile of the
distances, between observed and simulated pseudo datasets from earlier
steps, is considered as the threshold value.  For the SABC algorithm, we
used $\epsilon=500$ in analogy with the PMCABC one.  All of the other
parameters are left at the default value of the package.  To choose the
above tuning parameters we run multiple pilot runs to detect the parameter
values providing the most stable and the best convergence results of the ABC
approximate posterior distribution.  After this first step, we proceed to
the performance evaluation tasks described in the main text.

\end{itemize}
\end{appendix}

\end{document}